%% file: main.tex
\documentclass[10pt]{article}

\input{preamble}

\InputIfFileExists{numbers.tex}{}{C
  \typeout{^^JWARNING: numbers.tex not found. Run:^^JC
    python3 agent/eval/make_paper_tables.py^^J}}

\InputIfFileExists{figures/extra_numbers.tex}{}{C
  \typeout{^^JWARNING: figures/extra_numbers.tex not found. Run:^^JC
    python3 figures/make_figures.py^^J}}

\InputIfFileExists{ci_numbers.tex}{}{C
  \typeout{^^JWARNING: ci_numbers.tex not found. Run:^^JC
    python3 agent/eval/bootstrap_ci.py^^J}}

\title{MagTrace: Tracing magnetic order through the evolving scientific literature}

\author[1,2]{Mengnan Sun\thanks{Corresponding author: \href{mailto:msun29@stanford.edu}{msun29@stanford.edu}}}

\author[1]{Cheng Peng\thanks{Corresponding author: \href{mailto:cpeng18@stanford.edu}{cpeng18@stanford.edu}}}

\affil[1]{Stanford Institute for Materials and Energy Sciences, SLAC National Accelerator Laboratory, 2575 Sand Hill Road, Menlo Park, CA 94025, USA}
\affil[2]{Stanford University, Stanford, CA 94305, USA}

\date{}

\begin{document}
\maketitle
\begin{abstract}
\input{sections/00_abstract}
\end{abstract}

\input{sections/01_introduction}
\input{sections/02_methods}
\input{sections/03_results}
\input{sections/04_discussion}
\input{sections/05_conclusion}

\input{sections/06_backmatter}

\bibliography{refs}

\input{sections/07_appendix}

\end{document}

%% file: preamble.tex
\usepackage[T1]{fontenc}
\usepackage[utf8]{inputenc}
\usepackage{lmodern}
\usepackage{microtype}

\usepackage{amsmath,amssymb}
\usepackage{graphicx}
\usepackage{booktabs}
\usepackage{array}
\usepackage{colortbl}
\newcommand{\rowrule}{\arrayrulecolor{black!35}\specialrule{0.4pt}{1.5pt}{1.5pt}\arrayrulecolor{black}}
\usepackage{multirow}
\IfFileExists{siunitx.sty}{\usepackage{siunitx}}{%
  \typeout{^^JNOTE: siunitx not installed; units must be typeset by hand.^^J}}
\usepackage{xcolor}
\usepackage[margin=1in]{geometry}
\usepackage{authblk}
\usepackage[numbers,square,sort&compress]{natbib}
\usepackage[colorlinks=true,allcolors=blue!60!black]{hyperref}
\usepackage{enumitem}

\usepackage{caption}
\usepackage{abstract}

\newif\ifdraftmode
\draftmodetrue

\makeatletter
\newcommand{\resdef}[2]{\@namedef{papernum@#1}{#2}}
\newcommand{\val}[1]{%
  \@ifundefined{papernum@#1}%
    {\ifdraftmode
       {\color{red}\textbf{\textsf{\small[[NUM?~#1]]}}}%
     \else
       \GenericError{}{Undefined paper number: #1}{}{}%
     \fi}%
    {\@nameuse{papernum@#1}}}
\newcommand{\ci}[1]{\val{#1_ci}}
\makeatother

\newcommand{\kvec}{\ensuremath{\mathbf{k}}}

\newcommand{\Szero}{closed-book}
\newcommand{\Sone}{single search}
\newcommand{\Sseven}{full agent}

\newcommand{\Soneup}{Single search}
\newcommand{\Ssevenup}{Full agent}
\newcommand{\magndata}{\textsc{magndata}}
\newcommand{\tool}[1]{\texttt{#1}}

%% file: numbers.tex
\resdef{n_materials}{18}
\resdef{n_orders}{22}
\resdef{n_headline}{18}
\resdef{n_no_cell}{4}
\resdef{n_keys_with_msg}{2}
\resdef{n_stale_pairs}{3}
\resdef{n_repeats}{6}
\resdef{n_magndata_materials}{8}
\resdef{n_magndata_entries}{13}

\resdef{s0_composite}{0.548}
\resdef{s0_cwf1}{0.536}
\resdef{s0_specificity}{0.721}
\resdef{s0_top1}{0.417}
\resdef{s0_citations}{0.425}
\resdef{s1_composite}{0.676}
\resdef{s1_cwf1}{0.743}
\resdef{s1_specificity}{0.411}
\resdef{s1_top1}{0.778}
\resdef{s1_citations}{0.812}
\resdef{s7_composite}{0.839}
\resdef{s7_cwf1}{0.874}
\resdef{s7_specificity}{0.730}
\resdef{s7_top1}{0.889}
\resdef{s7_citations}{0.867}

\resdef{gate_unit_cell_delta_s0}{+0.005}
\resdef{gate_unit_cell_delta_s1}{+0.000}
\resdef{gate_unit_cell_delta_s7}{+0.001}
\resdef{gate_temperature_delta_s0}{+0.002}
\resdef{gate_temperature_delta_s1}{+0.000}
\resdef{gate_temperature_delta_s7}{+0.012}
\resdef{n_comparisons}{975}
\resdef{n_axis_verdicts}{123}
\resdef{simplex_grid_points}{1771}
\resdef{s0_ece}{0.266}
\resdef{s0_hiconf_conf}{87}
\resdef{s0_hiconf_correct_pct}{65}
\resdef{s1_ece}{0.235}
\resdef{s1_hiconf_conf}{88}
\resdef{s1_hiconf_correct_pct}{76}
\resdef{s7_ece}{0.135}
\resdef{s7_hiconf_conf}{89}
\resdef{s7_hiconf_correct_pct}{85}
\resdef{s0_zeros}{49}
\resdef{s0_abstentions}{6}
\resdef{s0_staleness}{0.33}
\resdef{s0_ladder_cov}{15}
\resdef{s1_zeros}{7}
\resdef{s1_abstentions}{6}
\resdef{s1_staleness}{0.69}
\resdef{s1_ladder_cov}{17}
\resdef{s7_zeros}{1}
\resdef{s7_staleness}{0.64}
\resdef{gap_s7_s0_composite}{0.291}
\resdef{gap_s7_s0_cwf1}{0.338}
\resdef{gap_s7_s1_composite}{0.163}
\resdef{gap_s7_s1_cwf1}{0.131}
\resdef{gap_s1_s0_composite}{0.128}
\resdef{gap_s1_s0_cwf1}{0.207}
\resdef{s0_contradicted_pct}{17}
\resdef{s0_citations_judged}{122}
\resdef{s0_citations_contradicted}{21}
\resdef{s1_contradicted_pct}{5}
\resdef{s1_citations_judged}{1274}
\resdef{s1_citations_contradicted}{65}
\resdef{s7_contradicted_pct}{7}
\resdef{s7_citations_judged}{550}
\resdef{s7_citations_contradicted}{36}
\resdef{s0_composite_prejudge}{0.543}
\resdef{s1_composite_prejudge}{0.676}
\resdef{s7_composite_prejudge}{0.833}

%% file: figures/extra_numbers.tex
\resdef{s7_stop_total}{108}
\resdef{s7_stop_verdict}{78}
\resdef{s7_stop_round_cap}{0}
\resdef{s7_stop_rule}{16}
\resdef{s7_stop_no_evidence}{14}
\resdef{s7_paper_cap_runs}{16}
\resdef{s7_conflict_rounds}{5}
\resdef{s7_repair_rounds}{0}
\resdef{s7_fallbacks}{0}
\resdef{s7_empty_nudges}{8}
\resdef{s7_salvaged}{0}
\resdef{s1_abstracts_median}{67}
\resdef{s7_early_stops}{14}
\resdef{s7_early_stop_materials}{4}
\resdef{s7_early_stops_laru}{5}
\resdef{s7_early_stops_ceir}{5}
\resdef{aczel_cited_earlier}{5}
\resdef{yre_top1_lost}{5}
\resdef{yre_repeats}{6}
\resdef{s0_confident_errors}{43}
\resdef{s0_orders_emitted}{158}
\resdef{s0_no_k_pct}{32}
\resdef{s1_orders_emitted}{372}
\resdef{s1_no_k_pct}{85}
\resdef{s7_orders_emitted}{229}
\resdef{s7_no_k_pct}{50}
\resdef{s0_citations_judged}{122}
\resdef{s1_citations_judged}{1274}
\resdef{s7_citations_judged}{550}
\resdef{s0_orders_unfound_pct}{25}
\resdef{s0_citations_unresolved}{42}
\resdef{s1_orders_unfound_pct}{2}
\resdef{s1_citations_unresolved}{0}
\resdef{s7_orders_unfound_pct}{0}
\resdef{n_unique_citations}{589}
\resdef{n_unique_citations_unresolved}{41}
\resdef{s0_unmatched}{95}
\resdef{s0_unmatched_supported}{12}
\resdef{s0_unmatched_offscope}{27}
\resdef{s0_unmatched_dup}{11}
\resdef{s0_fp_cost}{28.3}
\resdef{s1_unmatched}{255}
\resdef{s1_unmatched_supported}{244}
\resdef{s1_unmatched_offscope}{192}
\resdef{s1_unmatched_dup}{50}
\resdef{s1_fp_cost}{4.9}
\resdef{s7_unmatched}{106}
\resdef{s7_unmatched_supported}{94}
\resdef{s7_unmatched_offscope}{69}
\resdef{s7_unmatched_dup}{7}
\resdef{s7_fp_cost}{13.0}

%% file: ci_numbers.tex
\resdef{gap_s1_s0_citations_ci}{[0.236, 0.539]}
\resdef{gap_s1_s0_composite_ci}{[0.016, 0.255]}
\resdef{gap_s1_s0_contradicted_pct_ci}{[-21\%, -4\%]}
\resdef{gap_s1_s0_cwf1_ci}{[0.082, 0.362]}
\resdef{gap_s1_s0_ece_ci}{[-0.155, 0.053]}
\resdef{gap_s1_s0_hiconf_correct_pct_ci}{[-9\%, 43\%]}
\resdef{gap_s1_s0_no_k_pct_ci}{[33\%, 69\%]}
\resdef{gap_s1_s0_specificity_ci}{[-0.473, -0.137]}
\resdef{gap_s1_s0_top1_ci}{[0.139, 0.593]}
\resdef{gap_s7_s0_citations_ci}{[0.283, 0.610]}
\resdef{gap_s7_s0_composite_ci}{[0.171, 0.427]}
\resdef{gap_s7_s0_contradicted_pct_ci}{[-20\%, -2\%]}
\resdef{gap_s7_s0_cwf1_ci}{[0.194, 0.511]}
\resdef{gap_s7_s0_ece_ci}{[-0.211, -0.019]}
\resdef{gap_s7_s0_hiconf_correct_pct_ci}{[-2\%, 51\%]}
\resdef{gap_s7_s0_no_k_pct_ci}{[-4\%, 38\%]}
\resdef{gap_s7_s0_specificity_ci}{[-0.137, 0.138]}
\resdef{gap_s7_s0_top1_ci}{[0.287, 0.667]}
\resdef{gap_s7_s1_citations_ci}{[-0.014, 0.139]}
\resdef{gap_s7_s1_composite_ci}{[0.102, 0.229]}
\resdef{gap_s7_s1_contradicted_pct_ci}{[-1\%, 4\%]}
\resdef{gap_s7_s1_cwf1_ci}{[0.053, 0.217]}
\resdef{gap_s7_s1_ece_ci}{[-0.121, 0.019]}
\resdef{gap_s7_s1_hiconf_correct_pct_ci}{[-3\%, 23\%]}
\resdef{gap_s7_s1_no_k_pct_ci}{[-47\%, -22\%]}
\resdef{gap_s7_s1_specificity_ci}{[0.205, 0.428]}
\resdef{gap_s7_s1_top1_ci}{[0.009, 0.250]}
\resdef{s0_citations_ci}{[0.281, 0.564]}
\resdef{s0_composite_ci}{[0.429, 0.655]}
\resdef{s0_contradicted_pct_ci}{[9\%, 26\%]}
\resdef{s0_cwf1_ci}{[0.381, 0.662]}
\resdef{s0_ece_ci}{[0.196, 0.391]}
\resdef{s0_hiconf_conf_ci}{[86, 88]}
\resdef{s0_hiconf_correct_pct_ci}{[41\%, 87\%]}
\resdef{s0_no_k_pct_ci}{[19\%, 47\%]}
\resdef{s0_orders_unfound_pct_ci}{[14\%, 39\%]}
\resdef{s0_specificity_ci}{[0.612, 0.830]}
\resdef{s0_staleness_ci}{[0.00, 0.67]}
\resdef{s0_top1_ci}{[0.222, 0.611]}
\resdef{s1_citations_ci}{[0.754, 0.861]}
\resdef{s1_composite_ci}{[0.607, 0.740]}
\resdef{s1_contradicted_pct_ci}{[4\%, 6\%]}
\resdef{s1_cwf1_ci}{[0.664, 0.812]}
\resdef{s1_ece_ci}{[0.165, 0.316]}
\resdef{s1_hiconf_conf_ci}{[87, 89]}
\resdef{s1_hiconf_correct_pct_ci}{[59\%, 98\%]}
\resdef{s1_no_k_pct_ci}{[75\%, 93\%]}
\resdef{s1_orders_unfound_pct_ci}{[0\%, 6\%]}
\resdef{s1_specificity_ci}{[0.289, 0.539]}
\resdef{s1_staleness_ci}{[0.08, 1.00]}
\resdef{s1_top1_ci}{[0.602, 0.926]}
\resdef{s7_citations_ci}{[0.825, 0.909]}
\resdef{s7_composite_ci}{[0.798, 0.881]}
\resdef{s7_contradicted_pct_ci}{[4\%, 9\%]}
\resdef{s7_cwf1_ci}{[0.822, 0.923]}
\resdef{s7_ece_ci}{[0.126, 0.239]}
\resdef{s7_hiconf_conf_ci}{[88, 90]}
\resdef{s7_hiconf_correct_pct_ci}{[70\%, 98\%]}
\resdef{s7_no_k_pct_ci}{[35\%, 67\%]}
\resdef{s7_orders_unfound_pct_ci}{[0\%, 0\%]}
\resdef{s7_specificity_ci}{[0.621, 0.831]}
\resdef{s7_staleness_ci}{[0.42, 0.83]}
\resdef{s7_top1_ci}{[0.768, 0.981]}
\resdef{ci_resamples}{10{,}000}

%% file: sections/00_abstract.tex
Much of the research on magnetic materials starts from their magnetic order, but confirming it from the papers by hand can take hours, and curated databases can be incomplete and outdated. In this work, we build MagTrace, a language-model agent that, given only a material's name, searches four literature databases, reads open-access full text and reports the magnetic orders the current literature supports, each with a confidence score and citations a reader can check. We evaluate its performance on \val{n_materials} materials against an answer key compiled from published papers, compared with two other conditions that vary in level of retrieval. On a score from 0 to 1 that combines correctness, detail, whether the most confident order is current, and citation integrity, MagTrace scores \val{s7_composite}; the same underlying language model scores \val{s1_composite} given the abstracts from one literature search and \val{s0_composite} answering from memory. With a single search, the model finds the right kind of order but omits the propagation vector, which describes how the pattern repeats through the material's crystal structure, in \val{s1_no_k_pct}\% of its orders. From memory, the model often cites titles that lead to no indexed paper (for \val{s0_orders_unfound_pct}\% of the orders we could check), and when a cited paper exists and has an abstract, the abstract contradicts the claim for \val{s0_contradicted_pct}\% of those citations, against \val{s7_contradicted_pct}\% for MagTrace.

%% file: sections/01_introduction.tex
\section{Introduction}
\label{sec:intro}

A magnetic order describes where the magnetic moments in a material point and how that pattern repeats across its internal crystal structure, the regular arrangement of its atoms. It is mainly measured by neutron diffraction: because neutrons carry their own magnetic moment, they scatter off the moments in the material, and the resulting scattering pattern shows how those moments are arranged \citep{rodriguezcarvajal1993}. Because magnetic order shapes a material's electronic structure, symmetry, and macroscopic properties \citep{smejkal2022,bernevig2022}, accurate and up-to-date knowledge of this order is essential for interpreting experiments, predicting material behavior, and discovering new materials or new uses for existing materials \citep{xu2020}. For example, a first-principles calculation of a magnetic material has to start from an initial magnetic order, and it is important which magnetic order serves as the starting point: with the wrong starting magnetic order, a correct calculation could arrive at the wrong conclusion. 

The magnetic order of a material is recorded as a combination of an ordering type (for example, ferromagnetic, antiferromagnetic, ferrimagnetic, non-collinear) and a propagation vector $\mathbf{k}$, which describes how the pattern repeats from one crystallographic unit cell to the next. When the data allow, a symmetry label such as a magnetic (Shubnikov) space group or an irreducible representation of the crystal's space group can also be part of the description of a material's magnetic order \citep{perezmato2015}. In this paper, a \emph{magnetic order} refers to this whole combination, and the \emph{kind of order} refers to the ordering type alone. 

Nowadays, magnetic orders are also used as labels in the training datasets of machine-learning models for magnetic properties, and such datasets need magnetic orders for thousands of materials. For materials scientists, checking the magnetic order of a single material means manually finding all relevant papers, figuring out what magnetic order each paper describes, and deciding which to trust when findings disagree. This process might take hours to days per material, which puts a dataset of the size machine learning needs out of reach if one relies solely on manual data gathering. The labels in such datasets therefore often come from calculations or curated databases rather than manual checks against the literature \citep{merker2022}. 

Curated databases therefore become very important, since scientists query them to get the needed information about materials. However, each entry in a carefully curated database reflects the state of the literature at the time the entry was created, which can create two major problems. First, the databases are incomplete: new materials continue to be reported, and most magnetic materials have no magnetic order stored in any database. As of October 2026, \magndata{} \citep{gallego2016a,gallego2016b}, the main curated collection of measured magnetic orders, holds 2,447 entries, while the Northeast Materials Database, built by language models from published papers, lists 67,573 magnetic materials but records at most the kind of order of each material, not the full description of its magnetic order \citep{itani2025}. Of the \val{n_materials} materials in our evaluation set, picked to represent a wide range of materials, only \val{n_magndata_materials} have an entry in \magndata{}. Second, databases can be outdated: existing magnetic orders may be revised as new experimental evidence emerges. UO$_2$ is one example: its magnetic order was revised over several decades of experiments \citep{lander2020}. As a result, one material often has several entries. In \magndata{}, of the 1,558 materials whose entries give a plain chemical formula, 464 have two or more entries (Appendix~\ref{app:magndata}). Some of these describe different phases or conditions, while others describe the same phase as different papers reported it, sometimes decades apart. In our evaluation set, 3 of the materials with a \magndata{} entry have entries that do not all agree (\S\ref{sec:conclusion-magndata}). Scientists who find several entries have to go back to the papers to learn which one describes the material as it is understood today.

Earlier work has tried to supply magnetic data at a larger scale in three ways. The first computes magnetic orders instead of collecting measured ones. For example, \citet{horton2019} built a high-throughput workflow that predicts each material's ground-state collinear magnetic order. A calculated order is a prediction rather than an accurate measurement, and the workflow is limited to collinear orders. Calculations can also disagree with one another: when three large databases of calculated material properties are compared two at a time, they disagree on whether a material is magnetic at all for up to 15\% of the materials they have in common \citep{hegde2023}. The second way extracts data from papers automatically. Text-mining tools and language models pull property values out of research papers \citep{swain2016,polak2024}, and for magnetic materials they have been used to build databases of transition temperatures \citep{court2018,zhang2024gpt}, the largest of which is the Northeast Materials Database. However, these databases have no information beyond the kind of order. The third way trains machine-learning models on the curated record itself. For example, \citet{fahmy2026} trained classifiers on \magndata{} entries to predict whether a material's propagation vector is zero, and used them to flag more than 6,840 calculated entries that are probably labeled ferromagnetic by mistake. Such models learn from the database and treat its entries as correct, so they also inherit its gaps and its outdated entries. Therefore, none of these approaches replaces the curated record of measured orders, and none of them checks that record against the current literature.

Simply asking a language model for a material's magnetic order is not a reliable alternative either. Earlier work found that language models can cite papers that do not exist \citep{walters2023}. Although newer models are believed to have improved substantially, questions about their reliability remain. Language-model agents that read the scientific literature are a newer option. We know of no agent that reads the literature to report magnetic orders, so we compare our agent with the closest general ones, which are built for the same job as ours: searching the literature, reading the papers and backing each answer with citations. The comparison shows which parts of our task these agents already solve and which they do not. For example, PaperQA and its successor PaperQA2 search full-text papers and answer scientific questions with citations \citep{lala2023,skarlinski2024}, and PaperQA2 also flags contradictions between papers. OpenScholar does the same over 45 million open-access papers, and its citations are as accurate as those written by human experts \citep{asai2026}. Similar agents in chemistry and materials science run tools, experiments and simulations \citep{bran2024,boiko2023,chaudhari2026}, or, like the agent of \citet{ansari2024}, collect values reported in papers into datasets. The closest in purpose to ours is LitMOF, which checks the entries of a crystal-structure database against the papers they came from and repairs the entries that are wrong \citep{kim2025litmof}. However, these agents answer free-text questions, collect values without judging between them, or check a database entry against its own source paper. None of them is asked to return a magnetic order detailed enough to be checked against the papers, or to decide which of several published results is current. This decision is hard for language models in general: they often give answers based on medical conclusions that have since been revised \citep{vladika2025}, and search agents are often misled by conflicting evidence even after they have retrieved everything needed to answer \citep{hu2026trek}. The closest work on magnetic orders, LLaMP, asks language models for a material's magnetic ordering, but grades only the kind of order against calculated records \citep{chiang2025}. 

This paper presents MagTrace, a large language model agent that, given a material's name, decides which tools to use \citep{yao2023react}, such as a literature search or a paper reader, and returns a list of the magnetic orders found for the material, with a confidence score and the supporting papers for each magnetic order. It works the way materials scientists search the literature by hand. Our domain-specific agent reports each order with its propagation vector and the citations, is graded against the experimental literature, and must say which of several published orders is current. \textbf{Table~\ref{tab:prior}} sets the previous approaches beside ours. Magnetic order is a demanding test for this kind of agent, because papers write one order in many equivalent ways and published orders are sometimes revised. We measure how well MagTrace performs on an evaluation set of 18 materials, built from published papers, that covers different types of materials with different amounts of published evidence. 

\begin{table}[t]
\centering
\caption{What earlier approaches provide for a material's magnetic order, set beside this work. The first five rows are approaches to magnetic data. The next two are agents that read papers for a purpose similar to ours, but not for magnetic orders.}
\label{tab:prior}
\small
\begin{tabular}{>{\raggedright\arraybackslash}p{0.30\linewidth}>{\raggedright\arraybackslash}p{0.14\linewidth}>{\raggedright\arraybackslash}p{0.14\linewidth}>{\raggedright\arraybackslash}p{0.14\linewidth}>{\raggedright\arraybackslash}p{0.14\linewidth}}
\toprule
approach & measured orders & full order (type and $\mathbf{k}$) & current literature & picks the current order \\
\midrule
curated database, \magndata{} \citep{gallego2016a,gallego2016b} & yes & yes & no, fixed at entry & no \\
\rowrule
calculated orders \citep{horton2019} & no & collinear only & no & no \\
\rowrule
text-mined databases \citep{court2018,zhang2024gpt,itani2025} & measured data, not orders & no, only the kind and temperatures & no, fixed at extraction & no \\
\rowrule
classifiers trained on \magndata{} and calculated labels \citep{merker2022,fahmy2026} & learned from database entries & no, a predicted class & no & no \\
\rowrule
LLaMP \citep{chiang2025} & no & no, only the kind & not for orders & no \\
\midrule
literature agents, PaperQA2 and OpenScholar \citep{skarlinski2024,asai2026} & yes & no, free text & yes & no \\
\rowrule
database-checking agent, LitMOF \citep{kim2025litmof} & measured structures, not orders & not for magnetic orders & only each entry's source paper & no \\
\midrule
this work, MagTrace & yes & yes & yes & yes \\
\bottomrule
\end{tabular}
\end{table}

%% file: sections/02_methods.tex
\section{Methods}
\label{sec:methods}
\label{sec:task}

\begin{figure}[tb]
\centering
\includegraphics[width=\linewidth]{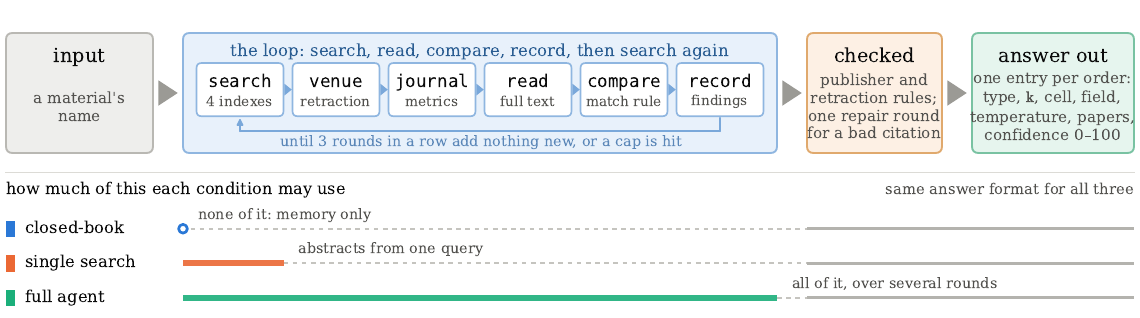}
\caption{The full agent, read left to right. With a material's name as the input, the loop
searches, vets, reads, compares and records until the model answers or code stops
it; code then applies the publisher and retraction rules, and the answer comes
out. The rows underneath show how much of the loop each condition of the evaluation (\S\ref{sec:rungs}) can use.}
\label{fig:agent}
\end{figure}

MagTrace mimics the process by which a scientist manually checks a material's magnetic order in the literature. It is given a material's name; a language model (\texttt{claude-sonnet-5}) then works in rounds, and in each round it can use a set of tools to search four literature databases for papers on the material, check whether a paper was retracted, read a paper's free full text, and compare different descriptions of a magnetic order. The answer is a list of the magnetic orders that the current literature supports; each order gives its ordering type, propagation vector, unit cell and the other details the papers state, together with a confidence score between 0 and 100 representing the agent's level of confidence that the order found is current and correct, and the citations behind every value (\textbf{Figure~\ref{fig:agent}}). The full output set is designed so that, once acquired, it provides all the necessary magnetic inputs for any density functional theory (DFT) software, given the crystal structure, to construct the correct unit cell and magnetic-order configuration. 

\subsection{Agent design}
\label{sec:agent}

The main loop of the agent has six tools, shown in \textbf{Table~\ref{tab:tools}}: two find and read papers, two check each paper's journal and retraction status, and two compare and keep track of the orders found. Appendix~\ref{app:agentdetail} describes each tool in full. Besides the material's name, the user can choose in the input which language model the agent runs on; all runs in the main results use \texttt{claude-sonnet-5}, and Appendix~\ref{app:usage} lists every input with examples.

The loop allows at most 80 papers, 16 rounds and 20 full-text reads per material; after 80 papers, searching closes, but reading and comparing can continue. Apart from these caps, code ends a run in two cases: before any order is recorded, a run ends if three consecutive searches find no new paper on the material (the \emph{early stop}); when the model has recorded an order, the run ends if three consecutive rounds find no new order (the \emph{stopping rule}). When the model's answer holds two contested magnetic orders that cannot both describe the material at ambient conditions, the loop gives the model one more search and one round of reading before it answers (the \emph{conflict round}). 

Before the final output is produced, code checks every paper the answer cites for its publisher and retraction status, and if the answer cites a paper the agent never retrieved or one that these rules exclude, the loop returns the answer to the model once so that it can replace those citations. Appendix~\ref{app:agentdetail} gives the exact rules and how often each one was used.

Each reported magnetic order carries a confidence score $p$ from $0$ to $100$. The model sets it by following a list of criteria: the most important factor is the level of verification or agreement between independent research groups on the magnetic orders (so a single unreplicated report should not receive a high confidence score); second-tier factors include publication recency and journal quality, while third-tier criteria are author affiliations, funding sources, citation counts and whether a paper is peer reviewed, which adjust the score only slightly (Appendix~\ref{app:agentdetail}). The model can also mark a magnetic order \emph{contested}, meaning the model judges that the current literature disputes that order. 

\begin{table}[t]
\centering
\caption{The six tools of the agent's main loop.}
\label{tab:tools}
\small
\begin{tabular}{>{\raggedright\arraybackslash}p{0.27\linewidth}>{\raggedright\arraybackslash}p{0.67\linewidth}}
\toprule
tool name & what it does \\
\midrule
\texttt{search\_paper} & Queries Semantic Scholar \citep{kinney2023}, Crossref \citep{hendricks2020}, OpenAlex \citep{priem2022} and arXiv together, removes duplicates, drops papers from suspicious publishers and papers not in English, and returns 10 papers per call. \\
\rowrule
\texttt{check\_venue\_status} & Reports a paper's publication type, publisher and retraction status. \\
\rowrule
\texttt{get\_journal\_quality} & Reports a journal's citation metrics. \\
\rowrule
\texttt{read\_paper} & Fetches only the free full text of the papers found and passes the model at most 8{,}000 characters of magnetism-related content per paper. \\
\rowrule
\texttt{compare\_orders} & Applies the match rule of \S\ref{sec:match} to two descriptions of an order, with the same judges the scorer uses. \\
\rowrule
\texttt{record\_candidate\_orders} & Lets the model write down the orders it has found, so that code can count new orders for the stopping rule. \\
\bottomrule
\end{tabular}
\end{table}

\subsection{Match rule}
\label{sec:match}

Papers often describe one magnetic order in different ways: they may write the propagation vector in different unit cells, choose different but symmetry-equivalent vectors, or write the symmetry label in different notations. The agent needs to know when two papers report the same order, so that it can count independent confirmations and notice disagreements, and we need the same decision to grade an answer against the answer key. One rule, the \emph{match rule}, makes this decision in both places: \texttt{compare\_orders} applies it inside the loop, and the scorer applies it when grading, so the agent is graded in evaluation by the standard it is told to use in implementation.

The general rule is that two orders match when they have the same ordering type and the same propagation vector expressed in the same unit cell. Vectors that differ only by a reciprocal-lattice vector (a shift that leaves the pattern unchanged), by an overall sign, or by the choice among symmetry-equivalent vectors still match. An order measured under an applied field, under pressure or in a doped sample is a different order from the ambient one and only matches another order with the same conditions. Appendix~\ref{app:match} gives the rule in full, including how it treats unit cells, transition temperatures and ferromagnets reported without a vector.

Some match decisions are too varied to make deterministically; these are made by a smaller model, \texttt{claude-haiku-4-5}. There are four such model judges that can be consulted during the agent loop and in scoring, but only when code finds a conflict and cannot decide on its own. The \emph{cell judge} decides whether two unit cells are the same cell written in different settings (equivalent choices of axes), the \emph{temperature judge} decides whether two conflicting transition temperatures describe the same phase, the \emph{condition judge} decides whether an order's stated conditions describe an applied field, pressure or a changed composition, and the \emph{adjudicator} decides whether two orders are the same, for example when two symmetry labels are written in different notations. A fifth, the \emph{citation judge}, is only called in evaluation, and decides whether a cited abstract supports the claim using the scoring criteria (\S\ref{sec:citations}). Appendix~\ref{app:judged} describes how each judge was checked against known answers.

\subsection{Evaluation}
\label{sec:eval}

To see how much searching and reading the literature adds, we compare MagTrace with two simpler approaches using the same language model, resulting in three conditions (\S\ref{sec:rungs}), on a set of 18 materials whose answer key we compiled from the published papers (\S\ref{sec:groundtruth}). Every answer is graded with one score made of four parts (\S\ref{sec:score}).

\subsubsection{Three conditions}
\label{sec:rungs}

The three conditions are defined to test the agent's performance: they differ only in what evidence reaches the model and what it is told about that evidence, so a difference in accuracy between them comes from the evidence. They are: \textbf{Closed-book}, meaning the model answers from memory with no access to paper copies or agentic tools. \textbf{Single search}, meaning one search using the query ``\textit{material} magnetic structure'' without any refinement: that query is sent to the OpenAlex database, asking for up to 80 papers, and the abstracts it returns (a median of \val{s1_abstracts_median} per material) go into the prompt in full; the model then answers once. \textbf{Full agent}, meaning MagTrace itself: in each \emph{round} the model can call the tools of \S\ref{sec:agent}, and it stops when it gives its answer or when code stops it.

Everything else is the same in all three conditions: the model (\texttt{claude-sonnet-5}) and its settings, the fields every answer must fill, the instructions for weighing evidence and assigning confidence, the match rule and the scorer (Appendix~\ref{app:conditions}). The only other differences are that closed-book, which has no way to look a paper up, cannot check a citation before giving it, and that only the full agent is told to exclude potentially predatory journals and retracted papers. Each material was run \val{n_repeats} times in each condition. We call one answer for one material in one condition a \emph{run}, and the $j$-th runs of every material and condition together make up \emph{repeat}~$j$.

\subsubsection{The evaluation set and answer key}
\label{sec:groundtruth}

The evaluation set is \val{n_materials} materials with \val{n_orders} orders between them: one current order for each material, a second current order for NiI$_2$, which has two ordered phases, and an older, \emph{superseded} order for each of the \val{n_stale_pairs} materials (UO$_2$, Na$_2$Co$_2$TeO$_6$ and CoPS$_3$) whose order later work revised. We compiled every value from the published papers and recorded it with its paper and page; none comes from a database entry (Appendix~\ref{app:groundtruth}). A field that no source states is left empty and is not scored, so a gap in sources does not lower an answer's score.

\subsubsection{Scoring}
\label{sec:score}

A ``key'' describes the ground-truth data about a material's magnetic order. The total score is a weighted sum of four parts, each between $0$ and $1$:
\begin{itemize}[leftmargin=*,itemsep=1pt,topsep=2pt]
\item \textbf{Correctness} ($45\%$): whether the reported orders are the ones in the key, weighted by the stated confidence score of the answer.
\item \textbf{Detail score} ($25\%$): how many of the details the key states a matched answer also states correctly.
\item \textbf{Top-1} ($15\%$): whether the order given the highest confidence matches a current order in the key.
\item \textbf{Citation integrity} ($15\%$): whether the cited papers exist, are eligible (not retracted, not from a suspicious publisher) and support the claim.
\end{itemize}
Appendix~\ref{app:score} gives the formulas for all four.

\paragraph{Correctness.} Each reported magnetic order is paired with at most one key order. A matched order earns full credit, and one that names only the kind of order (a \emph{type-only match}), such as ``antiferromagnetic'' with no propagation vector, earns half. Correctness combines how much of the key an answer found (recall) with how much of what it reported is right (precision) into a confidence-weighted $F_1$: every reported order that matches nothing lowers precision, and the more confidently it was stated, the more it costs. An unmatched order costs nothing when it restates an order already matched, or when it describes the material under conditions the key does not cover, such as high pressure or a monolayer, and a cited abstract supports it; we call this the \emph{other-conditions rule}.

\paragraph{Detail score.} For each matched order, the detail score checks four levels against the key: how finely the ordering type is named (for example, if the key states ``zigzag antiferromagnetic'', then an answer of just ``antiferromagnetic'' is not as fine), the propagation vector, the unit cell and a symmetry label, weighted $40\%$, $32\%$, $16\%$ and $12\%$. A level the key does not state is skipped. Because the detail score is computed only for matched orders, a condition that matches nothing on a material has no detail score there (Appendix~\ref{app:vocabulary}).

\paragraph{Citation integrity.}
\label{sec:citations}
Each citation is checked in two steps. First, we examine the paper's existence and its eligibility. Then, we ask the citation judge whether the paper's abstract supports the claim, contradicts it or does not address it; judging every citation against its abstract keeps the check equally strict for open-access and paywalled papers. A citation by DOI to a real paper whose abstract supports the claim earns full credit, while a citation to a paper that does not exist, is ineligible or is contradicted by its abstract sets the order's citation score to $0$ no matter what else the order cites (Appendix~\ref{app:citations}).

All scores are computed from the saved answers, with every judge replayed from its saved verdicts. We report the mean over the \val{n_repeats} repeats and the standard deviation of the \val{n_repeats} repeat totals. Each result in \S\ref{sec:results} also carries a 95\% confidence interval, found by resampling the materials, which shows how much the result could change on a different set of materials (Appendix~\ref{app:repro}).

%% file: sections/03_results.tex
\section{Results}
\label{sec:results}

Our results compare the three conditions in four ways: their total scores and the four parts that make up the total score (\S\ref{sec:results-main}), the papers each condition cites (\S\ref{sec:results-citations}), how correct each condition is on each material (\S\ref{sec:results-permaterial}), and how well each tells a current order from a superseded one and states confidence scores that match its accuracy (\S\ref{sec:results-staleness}). For the total scores and the gains between conditions, the numbers in brackets give the 95\% confidence interval, which shows how much the result could change on a different set of \val{n_headline} materials; \textbf{Table~\ref{tab:ci}} in Appendix~\ref{app:repro} gives the interval for every other number in this section.

\subsection{How the three conditions compare}
\label{sec:results-main}
\label{sec:results-cwf1}
\label{sec:results-specificity}
\label{sec:results-matchrule}
\label{sec:results-weights}

\begin{figure}[!t]
\centering
\includegraphics[width=0.823\linewidth]{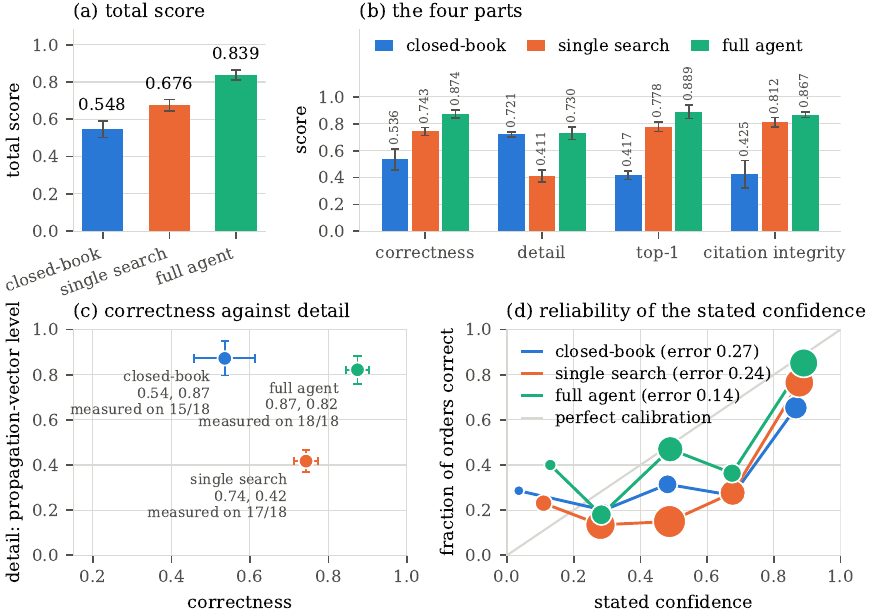}
\caption{The three conditions. (a) The total score and (b) its four parts; bars
are means over \val{n_repeats} repeats and whiskers one standard deviation across repeats. (c)
Correctness against the propagation-vector level of the detail score (the share of
matched orders whose propagation vector agrees with the key); each condition is
labeled with the number of materials on which its detail score could be computed. (d)
Calibration: stated confidence against the fraction of orders at that confidence that
are correct, with one point per confidence range of $20$; larger markers hold more
orders, and points below the diagonal mean the condition is overconfident. The
``error'' in the legend is the expected calibration error (\S\ref{sec:results-calibration}).}
\label{fig:conditions}
\label{fig:diagnostics}
\end{figure}

MagTrace scores highest of the three conditions. The total score rises as the model is given more evidence: it is \val{s0_composite} \ci{s0_composite} for closed-book, \val{s1_composite} \ci{s1_composite} for the single search and \val{s7_composite} \ci{s7_composite} for the full agent (\textbf{Figure~\ref{fig:conditions}a}). The score therefore rises in two steps of similar size: by \val{gap_s1_s0_composite} \ci{gap_s1_s0_composite} when the model goes from answering from memory to reading the abstracts of a single search, and by a further \val{gap_s7_s1_composite} \ci{gap_s7_s1_composite} when it goes from the single search to the full agent. Neither interval includes zero, so neither gain is likely to come from the particular materials we chose. The full agent also scores highest for all \val{simplex_grid_points} weighting schemes over the four score parts defined in \S\ref{sec:score} that we tested (every combination in steps of $5\%$), so its lead does not depend on the weights we chose (Appendix~\ref{app:score}).

The four parts show where each gain in the score comes from (\textbf{Figure~\ref{fig:conditions}b}). Correctness, the largest part of the score, rises in the same two steps as the total score: from \val{s0_cwf1} for closed-book to \val{s1_cwf1} for the single search and \val{s7_cwf1} for the full agent. Top-1 accuracy, whether the order given the highest confidence is a current order in the key, rises from \val{s0_top1} to \val{s1_top1} and \val{s7_top1}, and citation integrity rises from \val{s0_citations} to \val{s1_citations} and \val{s7_citations}. For both of these parts, most of the gain comes with the single search, and the full agent adds a smaller further gain, which for citation integrity is within the uncertainty (the interval on that gain is \ci{gap_s7_s1_citations}).

The detail score, which measures how many of the details in the key a matched order states correctly, ranks the conditions in a different order: the full agent scores \val{s7_specificity}, closed-book \val{s0_specificity} and the single search \val{s1_specificity}, so the full agent and closed-book are level (the interval on their difference is \ci{gap_s7_s0_specificity}). The single search therefore gives the least detail of the three, even though its answers are correct far more often than closed-book's. The difference lies in the propagation vector (Appendix~\ref{app:score}, \textbf{Table~\ref{tab:specificity}}). The single search leaves the vector out of \val{s1_no_k_pct}\% of its magnetic orders, against \val{s0_no_k_pct}\% for closed-book and \val{s7_no_k_pct}\% for the full agent: when the model reads only abstracts, it finds the right kind of order but does not state its vector. Closed-book's detail score is high for a different reason. A detail score can only be computed on a material where the condition matched at least one order, and closed-book did so on only \val{s0_ladder_cov} of the \val{n_headline} materials (against \val{s1_ladder_cov} for the single search and all \val{n_headline} for the full agent). Its detail score is therefore an average over the materials the model already knew well (\textbf{Figure~\ref{fig:conditions}c}).

\subsection{Citations}
\label{sec:results-citations}

\begin{figure}[!t]
\centering
\includegraphics[width=0.883\linewidth]{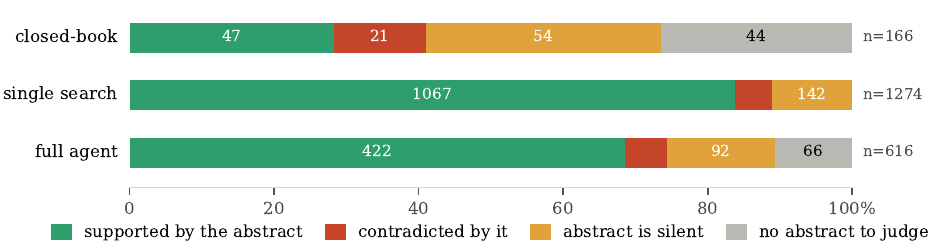}
\caption{What the citation judge found when it compared the abstract of each cited
paper with the claim it was cited for, by condition. Counts are citations, not magnetic orders.}
\label{fig:citoutcomes}
\end{figure}

Searching for papers makes citations far more reliable. Citation integrity, which checks whether the cited papers exist, are eligible and support the claim, is \val{s0_citations} for closed-book, against \val{s1_citations} for the single search and \val{s7_citations} for the full agent (\textbf{Figure~\ref{fig:conditions}b}). Closed-book's citations fail in two ways. Firstly, closed-book usually cites a paper by its title alone, and for \val{s0_orders_unfound_pct}\% of its checkable magnetic orders (orders with at least one citation that could be looked up) none of the cited titles leads to a paper found in the literature databases. Secondly, when the cited paper does exist and has an abstract, the abstract contradicts the claim for \val{s0_contradicted_pct}\% of closed-book's citations, against \val{s1_contradicted_pct}\% for the single search and \val{s7_contradicted_pct}\% for the full agent (\textbf{Figure~\ref{fig:citoutcomes}}). Because these papers are real, a check that only asks whether a cited paper exists would accept them. \textbf{Table~\ref{tab:citations}} in Appendix~\ref{app:citations} gives the same comparison per order instead of per citation.

\textbf{Figure~\ref{fig:citoutcomes}} breaks each condition's citations down by the citation judge's verdict. For the two conditions that search, most cited abstracts support the claim, and of the rest, more are abstracts that do not address the claim than abstracts that contradict it. Closed-book's citations look different: fewer than half are supported, many of the cited papers have no abstract the judge could read, and contradicted citations make up a larger part of the bar, \val{s0_citations_contradicted} of the \val{s0_citations_judged} citations with an abstract, against \val{s1_citations_contradicted} of \val{s1_citations_judged} for the single search and \val{s7_citations_contradicted} of \val{s7_citations_judged} for the full agent.

\subsection{Correctness on each material}
\label{sec:results-permaterial}
\label{sec:results-closedbook}

\begin{figure}[t]
\centering
\includegraphics[width=0.847\linewidth]{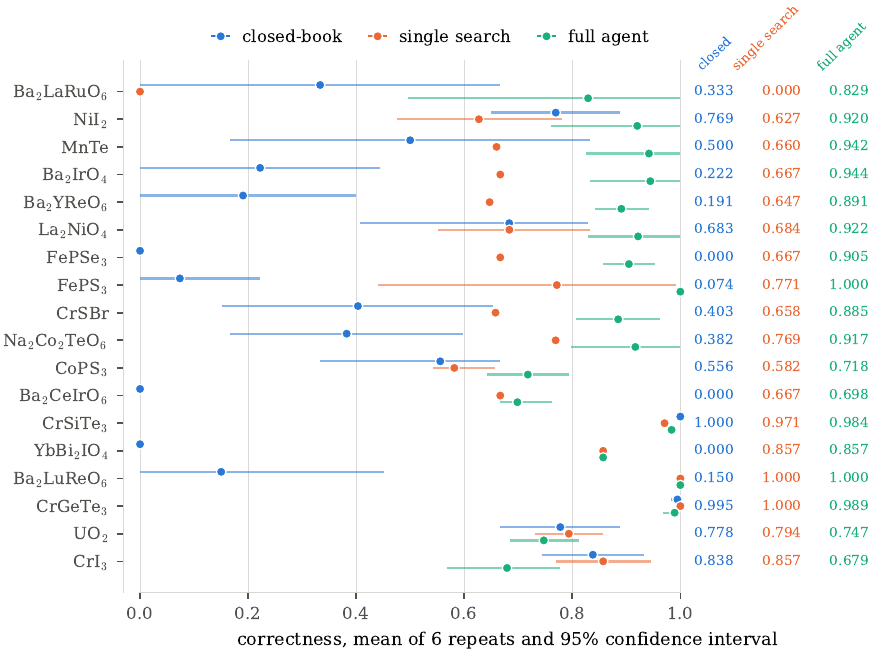}
\caption{Correctness on each material, sorted by how far the full agent's correctness
is above the single search's, with the largest difference at the top. Each point is
the mean over \val{n_repeats} repeats, and each bar its 95\% confidence interval from
resampling those repeats (Appendix~\ref{app:repro}).}
\label{fig:permaterial}
\end{figure}

The averages of correctness against detail hide large differences between materials for the closed-book and single-search conditions. On a single material, closed-book's correctness ranges from $0$ to $1$, and so does the single search's, while the full agent's correctness is above two thirds on every material (\textbf{Figure~\ref{fig:permaterial}}). Counting only materials where the 95\% confidence intervals do not overlap, the full agent has the highest correctness of the three conditions on 6 of the \val{n_headline} materials (Ba$_2$IrO$_4$, Ba$_2$YReO$_6$, CrSBr, FePS$_3$, FePSe$_3$ and Na$_2$Co$_2$TeO$_6$), and closed-book is highest on 1 material (CrSiTe$_3$). On the other 11 materials, the intervals of the highest condition and at least one other overlap, so we cannot decide which condition is highest.

In \textbf{Figure~\ref{fig:permaterial}}, each row is one material, with one point per condition and the three mean values printed at the right. The rows are sorted so that the materials on which the full agent gains most over the single search are at the top. The full agent's points sit toward the right in every row, while closed-book's are spread across the whole range. The gain over the single search is largest on Ba$_2$LaRuO$_6$, where the single search scores zero in every repeat, and the full agent's mean is below the single search's on only the 3 materials at the bottom, CrGeTe$_3$, UO$_2$ and CrI$_3$, where the intervals overlap.

Closed-book also fails completely most often, and it usually does so with high confidence. Out of 108 answers per condition (\val{n_headline} materials, \val{n_repeats} repeats each), closed-book scores exactly zero on \val{s0_zeros}, against \val{s1_zeros} for the single search and \val{s7_zeros} for the full agent. Only \val{s0_abstentions} of closed-book's zero scores are abstentions, in which the model declined to name an order; the other \val{s0_confident_errors} are confident errors, wrong answers stated with a confidence above zero. For example, on FePS$_3$ and FePSe$_3$, closed-book named the right ordering type, zigzag antiferromagnetic, in every repeat, but with confidence $55$ to $90$ it gave a propagation vector that differs from the one in the key. A confident error is dangerous because it gives the reader a wrong order with nothing to suggest that it needs checking. 

The two conditions that search also report many more orders than closed-book: \val{s1_orders_emitted} orders for the single search and \val{s7_orders_emitted} for the full agent, against \val{s0_orders_emitted} for closed-book. Most of the extra orders that match nothing in the key are published claims that the key does not include: the citation judge found an abstract that supports the claim for \val{s1_unmatched_supported} of the single search's \val{s1_unmatched} unmatched orders and for \val{s7_unmatched_supported} of the full agent's \val{s7_unmatched}, against \val{s0_unmatched_supported} of \val{s0_unmatched} for closed-book. Many of these extra orders describe the material under conditions other than the key's, such as high pressure, intercalation (atoms or molecules inserted between the layers) or a single layer of the material (a monolayer). Under the other-conditions rule, such an order does not lower the score when a cited abstract supports it. Appendix~\ref{app:permaterial} lists what the unmatched orders are, and describes 3 materials on which the stopping rule or our answer key shaped the outcome.

\subsection{Current against superseded orders, and confidence}
\label{sec:results-staleness}
\label{sec:results-calibration}

Telling the currently established magnetic order from an earlier, incompletely resolved one (a \emph{superseded} order) is especially important for keeping a database up to date, and 3 materials in our evaluation set test it: UO$_2$, Na$_2$Co$_2$TeO$_6$ and CoPS$_3$ each have both a current and a superseded order in the key. An answer scores $1$ when it gives the current order a higher confidence than the superseded one, $0.5$ when it reports the current order and leaves the superseded one out, and $0$ otherwise. The average scores are \val{s0_staleness} for closed-book, \val{s1_staleness} for the single search and \val{s7_staleness} for the full agent; with so few materials, the intervals are wide (\textbf{Table~\ref{tab:ci}}). On UO$_2$ every condition reports the current triple-\kvec{} order in every repeat, and on CoPS$_3$ the full agent reports the current order in every repeat, closed-book in 4 of the 6 and the single search in 2. Na$_2$Co$_2$TeO$_6$ is the hard case: the two conditions that search usually report two orders and mark them as contested, but the full agent gives the current order the higher confidence in only 2 answers, against all 6 for the single search (Appendix~\ref{app:permaterial}). With only 3 materials, this score is an observation rather than a reliable rate, and it is not part of the total score (Appendix~\ref{app:score}).

Because correctness charges a wrong order more the more confidently it is stated, the confidence scores need to be trustworthy. The expected calibration error measures this: it is the average gap between the confidence an answer states and how often orders stated at that confidence are actually correct, so lower is better. The error is \val{s0_ece} for closed-book, \val{s1_ece} for the single search and \val{s7_ece} for the full agent (\textbf{Figure~\ref{fig:conditions}d}); only the difference between closed-book and the full agent is clear of the uncertainty (\ci{gap_s7_s0_ece}). The same pattern appears for the most confident orders. When the model has a stated confidence of $80$ or above, it is about
equally confident in all three conditions: these orders have an average confidence score of
\val{s0_hiconf_conf} for closed-book, \val{s1_hiconf_conf} for the single search
and \val{s7_hiconf_conf} for the full agent. However, how often these high-confidence answers are right differs
across conditions: \val{s0_hiconf_correct_pct}\% of the closed-book orders, \val{s1_hiconf_correct_pct}\% of the single-search orders, and \val{s7_hiconf_correct_pct}\% of the full agent's match an
order in the key. With \val{n_headline} materials, these percentages are uncertain: the interval on the difference between closed-book and the full agent is \ci{gap_s7_s0_hiconf_correct_pct}, which includes zero. Still, the confidence number alone
does not tell a reader which answers to trust, and MagTrace's confident orders are the closest to their stated confidence.

\subsection{Other agent models}
\label{sec:results-models}
All results above are for \texttt{claude-sonnet-5}. Because the model is an input that the user chooses (Appendix~\ref{app:usage}), we checked that the full agent also runs correctly on other Claude models. We ran it once on CrI$_3$ with each of nine models (\texttt{claude-sonnet-5}, \texttt{claude-opus-5-5}, \texttt{claude-fable-5-1}, \texttt{claude-opus-5}, \texttt{claude-opus-4-8}, \texttt{claude-opus-4-7}, \texttt{claude-opus-4-6}, \texttt{claude-sonnet-4-6} and \texttt{claude-haiku-4-5}), with the tools returning results saved from one of this paper's runs so that every model saw the same evidence. This check used the six tools for at most three rounds, without the extra answer fields of Appendix~\ref{app:usage} and with \tool{compare\_orders} applying the fixed rule only. For all nine models, every request was accepted, no answer was cut off, and the final answer gave at least one order with an ordering type, a confidence and citations; three other models (\texttt{claude-sonnet-5-5}, \texttt{claude-haiku-5-5} and \texttt{claude-fable-5}) could not be tried because the model service we used does not offer them. This check shows that the agent runs on these models, not how well it does on them: of these, the evaluation of \S\ref{sec:eval} was also run on \texttt{claude-opus-5-5}, \texttt{claude-fable-5-1} and \texttt{claude-opus-5} (Appendix~\ref{app:models}), and on none of the others.

%% file: sections/04_discussion.tex
\section{Discussion}
\label{sec:discussion}
\label{sec:discussion-retrieval}
\label{sec:discussion-specificity}

\subsection{Three lessons learned from the three conditions}
Our research surfaced three important findings for anyone who needs a magnetic order they can trust, whether for their own work or to fill or correct a database.

\paragraph{Firstly, we cannot trust a model's memory.} Closed-book is not simply less accurate than the other two conditions; it is wrong with confidence. Of its \val{s0_zeros} answers that score zero, \val{s0_confident_errors} state a wrong order with a confidence above zero, and its orders stated at confidence $80$ or above are correct only \val{s0_hiconf_correct_pct}\% of the time, against \val{s7_hiconf_correct_pct}\% for MagTrace (although with \val{n_headline} materials, the interval on this difference includes zero). Closed-book's citations make this worse: for \val{s0_orders_unfound_pct}\% of closed-book's checkable magnetic orders with cited titles, the cited paper cannot be found in the literature databases. However, checking that a cited paper exists is not enough, because a real paper can still contradict the claim it is cited for. When the cited paper exists, its abstract contradicts closed-book's claim \val{s0_contradicted_pct}\% of the time, against \val{s7_contradicted_pct}\% for MagTrace. This is particularly concerning because a reader may be more likely to accept a claim, without checking whether the paper actually supports the stated magnetic order, when it comes with a real reference than when it comes with none. 

\paragraph{Secondly, a simple search in scientific databases finds the type of order but not its details.} The single search is right far more often than closed-book, yet it gives the least detail of the three, leaving the propagation vector out of \val{s1_no_k_pct}\% of its orders. An answer that says ``antiferromagnetic'' without a vector cannot be tied to one published order, used to start a calculation, or checked against the vector a paper reports. A reader shown only the correctness score would conclude that a single search is a clear improvement over answering from memory, while a reader shown only the detail score would conclude the opposite. This is why we report the four parts of the score alongside the total (Appendix~\ref{app:score}).

\paragraph{Thirdly, each kind of evidence improves a different part of the answer.} Correctness rises in two steps: one when the model is given the single search's abstracts, and one when it uses the full agent's repeated searching and reading. Most of the gain in citation integrity comes from the single search, which raises it from \val{s0_citations} to \val{s1_citations}; the full agent's further rise, to \val{s7_citations}, is within the uncertainty. The gain in detail comes from the full agent, which states the propagation vector far more often than the single search and brings the detail score back to closed-book's level. Therefore, a system built for this task needs retrieval for citations a reader can check, and the full loop of searching and reading for answers that are both right and specific.

\subsection{Potential use cases of MagTrace}

MagTrace is ready for two near-term uses, described below, and its design leaves room to grow in three directions. A larger answer key, with more materials whose order was later revised (only \val{n_stale_pairs} of ours were) and with sources that state the unit cell for the \val{n_no_cell} materials where ours do not, will measure more precisely how well the agent tells a current order from a superseded one. The full agent's confidence scores are better calibrated than closed-book's (expected calibration error \val{s7_ece}, against \val{s0_ece}), which is what a ranking of database entries by the agent's confidence that they disagree with the literature requires; such a ranking would point a curator to the entries that most need checking, and testing it is a natural next step. Finally, the score is built to be extended to what matters for a user's own question, such as transition temperatures or the temperature range of each phase, which the agent already reports: adding a new criterion only requires declaring a new answer field, a rule for comparing it and its weight in the score (Appendix~\ref{app:score}).

\paragraph{Cross-checking a database}
\label{sec:conclusion-magndata}

\magndata{} holds \val{n_magndata_entries} entries for \val{n_magndata_materials} of our \val{n_materials} materials, and we compared the order each entry claims with the answer key and with the full agent's answers (Appendix~\ref{app:magndata}). For 5 of these materials, the entries agree with the key's current order on the propagation vector, and the full agent's most confident order matches that order in every repeat. The other 3 have multiple database entries that do not all agree, so the database alone cannot tell a reader which magnetic order describes the material as it is widely accepted today. Of the 4 entries for Na$_2$Co$_2$TeO$_6$, 2 record the single-\kvec{} zigzag order that our key's sources describe as superseded; and the 2 MnTe entries, like the 2 La$_2$NiO$_4$ entries, agree with the key on the propagation vector but give different magnetic space groups. A database can therefore hold two magnetic orders of one material side by side without saying which is superseded, while MagTrace can point to the papers that decide it.

\paragraph{Finding open questions, and next steps}
\label{sec:conclusion-debates}
\label{sec:discussion-use}
\label{sec:future}
\label{sec:conclusion-extend}

An agent that reads the literature on a material to establish its magnetic order also sees where that literature disagrees with itself, and a disagreement between two careful measurements is a research question rather than an error in the record. MagTrace already turns up such cases. On Na$_2$Co$_2$TeO$_6$ it reports the single-\kvec{} and triple-\kvec{} descriptions side by side and marks them contested (Appendix~\ref{app:limitations}); on Ba$_2$YReO$_6$ it reports a newer transition temperature than our key holds; and on most materials it reports published orders under pressure, in thin samples or from calculations that our key does not hold (Appendix~\ref{app:permaterial}). For every order, each answer already records the fields on which the papers disagree, with each reading and the paper it comes from; gathered across materials, these records would become a list of open questions.\\

\subsection{Limitations}
\label{sec:limitations}
\label{sec:discussion-limits}

Our evaluation has six main limitations, listed below; Appendix~\ref{app:limitations} gives more detail on three of them.

\paragraph{Telling a current order from a superseded one rests on 3 materials.} Only UO$_2$, Na$_2$Co$_2$TeO$_6$ and CoPS$_3$ have both a current and a superseded order in the key, so this capability, which matters most for keeping a database current, is the one we measure least well.

\paragraph{The answer key is partially incomplete.} No source states a unit cell for \val{n_no_cell} materials, and only \val{n_keys_with_msg} orders carry a magnetic space group, so those checks are skipped. The key holds one order for most materials, so when an answer reports a published phase that the key does not list, that order matches nothing in the key and lowers the answer's correctness score (more so the more confidently it is stated), unless the other-conditions rule applies. On Ba$_2$YReO$_6$ the key also disagrees with a newer study on the transition temperature (Appendix~\ref{app:permaterial}), which lowers the full agent's score in the top-1 field in most repeats.

\paragraph{The scorer and its judges.} Several scoring rules were set after the answers had been read; Appendix~\ref{app:postrun} lists each with its effect, and the conditions rank in the same order without any of them. The judges were checked only on conflicts from our own runs, so how well they do on other materials is not known, and the citation judge was wrong on 8 of 31 hard cases labeled by hand, most often by accepting a specific magnetic pattern when the abstract describes it only in more general terms (Appendix~\ref{app:judged}).

\paragraph{The citation check.} Faithfulness is judged against abstracts, so a claim that only the full text contradicts can pass, and a paper without an abstract in the bibliographic databases is not judged at all. A citation given only by a title that resembles a paper in the databases without clearly matching it is left out of the check: \val{s0_citations_unresolved} of closed-book's citations and \val{s1_citations_unresolved} of the single search's (Appendix~\ref{app:citations}).

\paragraph{The early stop.} Ending a search after 3 searches in a row find nothing new on the material saves effort on materials with little literature, but the early stop can end the search before a decisive paper appears, especially one whose title does not name the material, as on Ba$_2$CeIrO$_6$ (Appendix~\ref{app:permaterial}).

\paragraph{What the comparison separates.} Every condition in the main results ran on one model, \texttt{claude-sonnet-5}; the same experiment on three newer models, run under the same conditions, gives the full agent the highest mean score with every model, but with two of them its lead over closed-book is within the uncertainty, and closed-book's mean is above the single search's (Appendix~\ref{app:models}). The full agent also differs from the single search in three ways at once (a search over four databases, many searches instead of one, and full-text reading), so we cannot say how much of its gain each one supplies. Score differences below about $0.05$ between repeated runs should be read as noise (Appendix~\ref{app:repro}).

\label{sec:bodyend}

%% file: sections/05_conclusion.tex
\section{Conclusion}
\label{sec:conclusion}

We graded MagTrace on \val{n_materials} testing materials against an answer key taken from the published papers. MagTrace searches four literature databases repeatedly and reads open-access full text, and it scores \val{s7_composite}, against \val{s1_composite} for a simple search over abstracts and \val{s0_composite} for the same model answering from memory. It comes first under every weighting of the score's four parts that we tried (Appendix~\ref{app:score}). A single search brings most of the gain in citation integrity, the full agent restores the detail that the single search loses, and both add to correctness. A model answering from memory often cites titles that lead to no real paper, and when the paper does exist its abstract contradicts the claim more often than for either condition that searches.

%% file: sections/06_backmatter.tex
\section*{Data availability}
\label{sec:data}
The answer key with the paper and page behind every value, the saved answers of
every run, the saved verdicts of every judge and the citation lookups are part of
the code repository below, so every number in this paper can be recomputed
offline. The papers the answer key is drawn from are copyrighted and are cited
rather than redistributed.

\section*{Code availability}
\label{sec:code}
The agent, the scorer and the scripts that generate every table and figure are
in a GitHub repository that will be made public on publication, together with a
tag for the revision that produced the results reported here.

\section*{Acknowledgments}

The work at Stanford University and SLAC National Accelerator Laboratory was supported by the U.S. Department of Energy, Office of Basic Energy Sciences, Division of Materials Sciences and Engineering, award number DE-SCL0000172. This research used resources of the National Energy Research Scientific Computing Center, a DOE Office of Science User Facility supported by the Office of Science of the U.S. Department of Energy under Contract No.~DE-AC02-05CH11231 using NERSC award BES-ERCAP0034968. M. Sun was supported by the Summer Internship Program at the Linac Coherent Light Source (LCLS) at the SLAC National Accelerator Laboratory, during the summer internship period. SLAC National Accelerator Laboratory is supported by the U.S. Department of Energy, Office of Science, Office of Basic Energy Sciences under Contract No. DE-AC0276SF00515. The system studied in this paper is built on Anthropic's \texttt{claude-sonnet-5} and \texttt{claude-haiku-4-5} models, as described in the Methods. Separately, AI tools were used to help write the code and to revise the language of this manuscript, under the authors' direction. The authors take full responsibility for the content of this work.

\section*{Author contributions}
M.S. designed and built the agent and the evaluation, compiled the answer key, ran the experiments, analyzed the results and wrote the manuscript. C.P. conceived the project, supplied source papers and propagation vectors for the answer key, reviewed the key, and advised on the framing of the work. Both authors reviewed the manuscript.

\section*{Competing interests}
The authors declare no competing interests.

%% file: sections/07_appendix.tex
\appendix

\section{The database cross-check}
\label{app:magndata}

Table~\ref{tab:magndata} sets each \magndata{} entry's claim beside our answer
key and the full agent's answers. Propagation vectors are compared only where
the entry and the key use the same parent setting; an entry is labeled not
comparable when the key states nothing it can be compared with, or when it
describes a different polymorph. The entries' values are their claims, read from
the entries themselves, and play no part in the answer key or in grading.

\paragraph{The whole database.} The counts in the introduction come from the
full \magndata{} listing, fetched once on 2026-10-17: 2{,}278 commensurate and
169 incommensurate entries (the listing's two sections), 2{,}447 in all. Grouping the entries by reduced
chemical composition, after setting aside 169 whose labels give no plain
formula (most write groups in parentheses or brackets; others are non-stoichiometric or carry a phase name), leaves 1{,}558 materials, of which 464 have two or more entries, 1{,}184 entries between them.

\begin{table}[t]
\centering
\caption{The \magndata{} entries for our materials, set beside the answer key
and the full agent. ``Agent'' summarizes the full agent's most confident order
over \val{n_repeats} repeats. A citation beside an entry number is the paper the entry cites; the claim is the entry's own.}
\label{tab:magndata}
\small
\begin{tabular}{p{0.15\linewidth}p{0.29\linewidth}p{0.47\linewidth}}
\toprule
material, entry & entry's claim & against the key and the agent \\
\midrule
UO$_2$, 3.2 & triple-\kvec{}, $\langle100\rangle$ star, $Pn\bar{3}m'$ &
agrees with the current order; the agent's top order matches its triple-\kvec{} order in every repeat, but names $Pa\bar{3}$ rather than $Pn\bar{3}m'$ in 4 of 6 \\
\rowrule
FePS$_3$, 1.183 & $\kvec=(0,1,\tfrac12)$ in $C2/m$ & agrees; agent agrees in
every repeat \\
\rowrule
FePSe$_3$, 1.210 & $\kvec=(\tfrac12,0,\tfrac12)$ in $R\bar{3}$ & agrees; agent
agrees in every repeat \\
\rowrule
CrSBr, 1.802 & $\kvec=(0,0,\tfrac12)$ in $Pmmn$ & agrees; agent agrees in every
repeat \\
\rowrule
CoPS$_3$, 1.264 & $\kvec=(0,1,0)$ in $C2/m$, $P_C2_1/m$ & agrees with the current order, from the same paper as the key; the agent names $P_C2_1/m$ in 5 of 6 repeats \\
\rowrule
Na$_2$Co$_2$TeO$_6$, 1.184 and 1.645 & single-\kvec{} $(\tfrac12,0,0)$,
$P_C2_12_12_1$ & the key's superseded zigzag order \\
\rowrule
Na$_2$Co$_2$TeO$_6$, 3.29 & triple-\kvec{}, $P6_32'2'$ & agrees with the current
order \\
\rowrule
Na$_2$Co$_2$TeO$_6$, 1.771 & parent $C2/m$, $\kvec=(\tfrac12,\tfrac12,0)$ & a
different (monoclinic) polymorph; not comparable \\
\rowrule
MnTe, 0.800 \citep{kunitomi1964} & $\kvec=0$, $Cmcm$ & agrees on \kvec{}; the key states no
magnetic space group \\
\rowrule
MnTe, 0.1155 \citep{liu2026mnte} & $\kvec=0$, $Cm'c'm$ & agrees on \kvec{}; the agent names
$Cm'c'm$ in 3 of 6 repeats \\
\rowrule
La$_2$NiO$_4$, 0.45 and 1.42 & $Pc'c'n$, $\kvec=0$; $P_Cmna$, $\kvec=(1,0,0)$ in
$Bmeb$ & both agree on \kvec{}: the key's $(\tfrac12,\tfrac12,0)$ is $(1,0,0)$ in
$Bmeb$ and a lattice vector, so $\kvec=0$, in the $P4_2/ncm$ cell of 0.45; the key
gives an irreducible representation, not a magnetic space group; the agent gives
the key's in-plane $(\tfrac12,\tfrac12)$ in every repeat \\
\bottomrule
\end{tabular}
\end{table}

\section{The full agent in detail}
\label{app:agentdetail}

\paragraph{Model.} All conditions run on \texttt{claude-sonnet-5}, with a cap of
20{,}000 tokens per reply and no sampling parameter sent. The judges, and the
extraction step inside the full-text reader, run on \texttt{claude-haiku-4-5}.

\paragraph{Tools.} 
\begin{description}[leftmargin=*,itemsep=2pt]
\item[\tool{search\_paper}] queries OpenAlex, Semantic Scholar, Crossref and
arXiv, merges and de-duplicates the results, drops papers from potentially predatory journals and non-English papers,
ranks papers that name the material first, and returns up to 10 papers it has not
shown before per call, with their metadata and abstracts. At most five papers in a row that do not name the
material are shown. When a query finds fewer than 3 papers, the tool repeats it with other spellings of the formula. A paper names the material when its title or abstract gives the name or formula in any common typography, including formulas written with spaces, subscripts or markup (``Ba$_2$IrO$_4$'', ``CoPS$_3$'').
\item[\tool{check\_venue\_status}] returns a paper's publication type, whether it
is peer reviewed, and whether it has been retracted or is from a journal of questionable credibility.
\item[\tool{get\_journal\_quality}] returns journal-level signals from OpenAlex,
with the SCImago quartile as a secondary signal.
\item[\tool{read\_paper}] fetches full text from free sources only (OpenAlex's
open-access copy, Semantic Scholar, CORE when an access key is set, Unpaywall, a repository page, and
arXiv last, so that a version of record is preferred to a preprint), reduces it
to at most 8{,}000 characters of magnetism-related content, and returns the
abstract when no free full text exists.
\item[\tool{compare\_orders}] applies the match rule of
Appendix~\ref{app:match} to two descriptions, through the same code and judges
as the scorer, with the material's crystal system and lattice centering supplied
by code.
\item[\tool{record\_candidate\_orders}] stores the orders the agent has found;
an order is new if, by the match rule without the judges and without its stated
conditions, it matches none already stored. The same order reported under a new
condition, such as high pressure, is therefore not new.
\end{description}

\paragraph{The loop.} 
Each material is capped at 80 papers, 16 rounds, 20 successful full-text reads
and 50 read attempts. Searching closes once 80 papers have been seen (in \val{s7_paper_cap_runs} of
the \val{s7_stop_total} runs), but reading and comparing continue. Of the \val{s7_stop_total} full-agent runs, \val{s7_stop_verdict} ended with the model giving its answer, \val{s7_stop_rule} after three rounds without a new order, \val{s7_stop_no_evidence} at the early stop, and \val{s7_stop_round_cap} at the 16-round cap. Once an order has been recorded, a round that adds no new order counts toward the stopping rule whatever the model did in it. Before any order is recorded, only a round that searched can count toward the early stop, and only if it found no new paper naming the material, read no full text and recorded nothing; rounds spent vetting or comparing leave the count unchanged. After either kind of round the model is told how many more such rounds it has before it must answer.
The conflict round of \S\ref{sec:agent} needs two orders in the model's answer that it marks contested, that code finds to be different orders, and that code confirms are under the same plain conditions: both experimental, on the bulk crystal of the standard structural form, at ambient pressure, without a field, and neither limited to a window of temperature. The model then gets one search, which must aim at settling the dispute, and one round in which it can read and compare but not search; the conflict round is not given when fewer than two rounds remain before the round cap or once the run has reached the paper cap, and it happened in \val{s7_conflict_rounds} runs.

\paragraph{Checks on the answer.} Code checks every paper the answer cites
for being from a journal flagged as potentially unreliable and for retraction status; in this experiment no paper was excluded on either ground.
If the answer cites a paper the agent never retrieved or an excluded paper, the loop returns it once
with those citations named and asks for them to be replaced; no answer in this experiment needed it (\val{s7_repair_rounds} of \val{s7_stop_total}). If the model's final reply contains no readable answer, the loop asks
once more, which happened in \val{s7_empty_nudges} runs, and every one of those runs then gave a readable answer. If that also failed, the loop would keep the last complete answer the model gave in the run or, if there was none, ask once more for an answer from the evidence already gathered; neither was needed (\val{s7_fallbacks} and \val{s7_salvaged} runs).

\paragraph{Confidence.} The confidence rubric ranks its factors in tiers. The
top factor is reproducibility: how consistently independent research groups
report the same order, judged with the match rule, so a single unreplicated report
should not reach high confidence. Publication date and journal quality come next,
and affiliations, funding, citation counts and whether a paper is peer reviewed adjust the
score at the margin. When newer work revises a well-replicated order under the
same conditions, the revision takes precedence. Papers from a journal flagged as potentially unreliable and retracted papers are
excluded outright.

\section{Running the agent: inputs and examples}
\label{app:usage}

The agent is called from Python with \texttt{run\_agent} in the code
repository's \texttt{agent/agent\_loop.py}. The only required input is the
material's name or formula; everything else has a default, and the defaults
reproduce the agent evaluated in this paper except where
\textbf{Table~\ref{tab:usage}} says otherwise. The language model is one of the inputs: the user can
run the agent on any Claude model by naming it, and changing the model leaves the
prompt, the tools and the caps unchanged. The judges and the extraction step inside \tool{read\_paper} run on
\texttt{claude-haiku-4-5}, whatever model the agent uses.

\begin{table}[htbp]
\centering
\caption{The inputs of \texttt{run\_agent}.}
\label{tab:usage}
\footnotesize
\begin{tabular}{>{\raggedright\arraybackslash}p{0.21\linewidth}>{\raggedright\arraybackslash}p{0.22\linewidth}>{\raggedright\arraybackslash}p{0.49\linewidth}}
\toprule
input & default & what it sets \\
\midrule
\texttt{material} (text) & required & The material's name or formula, e.g.\ \texttt{"CrI3"}. \\
\rowrule
\texttt{model} (text) & \texttt{\$ANTHROPIC\_MODEL} if set, else \texttt{claude-sonnet-5} & The language model of the agent: any Claude model name. One model serves the whole run. \\
\rowrule
\texttt{tools} (list) & the six tools of \textbf{Table~\ref{tab:tools}} & Which tools the agent may call. \tool{get\_magndata\_reference}, which looks up the paper a MAGNDATA entry cites, can be added but was off in this paper's runs. \\
\rowrule
\texttt{claimed\_orders} (list) & none & Order(s) to check against the literature (verification). With none, the agent reports what the literature supports, as in this paper. \\
\rowrule
\texttt{magndata\_entry} (text) & none & A MAGNDATA entry id, used only when \tool{get\_magndata\_reference} is enabled. \\
\rowrule
\texttt{max\_papers}, \texttt{max\_rounds} (numbers) & 80, 16 & The paper and round caps of Appendix~\ref{app:agentdetail}. \\
\rowrule
\texttt{criteria} (list) & none & Extra answer fields: any of \texttt{transition\_temperature}, \texttt{temperature\_range}, \texttt{competing\_states}, \texttt{dimensionality}, \texttt{field\_dependence}. This paper's runs asked for \texttt{transition\_temperature}, \texttt{temperature\_range} and \texttt{dimensionality}, in every condition; they are not scored. \\
\rowrule
\texttt{disagreements} (yes/no) & no & Adds a list of the fields on which the papers disagree, with each reading and its source. On in this paper's runs, in every condition. \\
\rowrule
\texttt{compare\_judges} & none & The judges \tool{compare\_orders} consults. With none it applies the fixed rule only; this paper's runs passed the scorer's judges. \\
\rowrule
\texttt{budget} & none & A token counter for the run. \\
\rowrule
\texttt{checkpoint\_path} (file) & none & Where to save the run after every round, so a crash loses at most one round. \\
\rowrule
\texttt{top\_p} (number) & not sent & A sampling setting, accepted only by models that still take one (not by \texttt{claude-sonnet-5}). \\
\rowrule
\texttt{verbose} (yes/no) & no & Prints each round. \\
\bottomrule
\end{tabular}
\end{table}

The answer is a dictionary with three parts: \texttt{orders}, the magnetic orders
with the fields of the output contract (Appendix~\ref{app:conditions});
\texttt{provenance}, which records the rounds, tool calls and papers seen and why
the run stopped; and \texttt{transcript}, the full exchange with the model. The model can also be chosen without changing the code, through the
environment variable \texttt{ANTHROPIC\_MODEL} or the \texttt{-{}-model} option of
\texttt{run\_ablation.py}. Example inputs:
{\small
\begin{verbatim}
from agent.agent_loop import run_agent

# claude-sonnet-5 and the six tools of this paper.
answer = run_agent("CrI3")

# The answer fields of this paper's runs.
answer = run_agent("CrI3", criteria=["transition_temperature",
                   "temperature_range", "dimensionality"],
                   disagreements=True)

# The same agent on another model.
answer = run_agent("CrI3", model="claude-opus-4-8")

# Another model, extra answer fields, and a checkpoint file.
answer = run_agent("NiI2", model="claude-haiku-4-5",
                   criteria=["transition_temperature"],
                   checkpoint_path="NiI2_checkpoint.json")

# Check a claimed order instead of discovering one.
answer = run_agent("MnTe", claimed_orders=[
    {"magnetic_order": "A-type antiferromagnetic",
     "propagation_vector": "(0, 0, 0)"}])

answer["orders"]       # the magnetic orders found
answer["provenance"]   # rounds, tool calls, papers, why it stopped
\end{verbatim}
}
\noindent and from the command line, for a material in the evaluation set:
{\small
\begin{verbatim}
python3 agent/eval/run_ablation.py --rungs S7 --materials CrI3 \
    --s7-tools full --model claude-opus-4-8 --repeats 1
\end{verbatim}
}

\section{The order-comparison rule in full}
\label{app:match}
Every condition is given the rule as text, and the rule is implemented once in
code, which both the scorer and the full agent's compare tool call. The code also
fixes three details that the text leaves to judgment: the temperature tolerance,
the moment-direction check and the treatment of doped samples. Two
descriptions are the same order when every one of the following holds. The rule
uses these terms.
\begin{description}[leftmargin=*,itemsep=1pt,font=\normalfont\itshape]
\item[Unit cell and setting.] The repeating box of the crystal. One lattice can be
described by several cells, called settings, and a cell may be \emph{centered}
(C, I, F or R), with extra lattice points inside it.
\item[Propagation vector $\kvec$.] How the magnetic pattern repeats from one unit
cell to the next, written in units of the reciprocal lattice of a chosen cell;
$\kvec=(0,0,0)$ means every cell carries the same moments.
\item[Star of $\kvec$.] The vectors related to $\kvec$ by the crystal's symmetry. A
\emph{single-}$\kvec$ order uses one of them, a \emph{multi-}$\kvec$ order several at
once.
\item[Symmetry labels.] A magnetic space group, the symmetry of the crystal
together with its moments, written in BNS notation with a catalog number; or
an irreducible representation, which labels how the moments transform under
the crystal's symmetry.
\end{description}

\paragraph{Ordering type and propagation vector.} They name the same ordering
type and the same propagation vector expressed in the same unit cell. Within one
cell, vectors that differ by a whole reciprocal-lattice step (with the steps a
centered cell allows), by an overall sign, or by the choice among the
symmetry-equivalent vectors of a star are the same. Two modulation types
(helical, cycloidal, longitudinal sinusoidal) that differ are different orders. An order
in which several symmetry-equivalent vectors order at once (multi-\kvec{}) is a
different order from one in which a single vector orders, and a stated number of
vectors (two against three) must agree; an answer that states no number asserts
none. When one side states no vector, the two match on type alone, unless both
name a specific pattern: two patterns that never describe the same order (zigzag
against N\'eel) are different orders, and other differing patterns go to the
adjudicator. A match on type alone earns half credit when the answer gave no vector where the key has one, and full credit when the key itself has none (\S\ref{sec:score}).

\paragraph{Unit cell.} A propagation vector counts steps of a chosen cell, so a
vector is compared only within one cell. Two standard descriptions of one lattice
(the hexagonal and rhombohedral settings of a rhombohedrally centered group, or the
primitive and conventional cells of one space group) are the same cell; a cell
that is not a setting of the other, a supercell, or the cell of a different
structural polymorph is a different cell. When code finds a conflict between two
stated cells and the lattice parameters do not settle it, the cell judge decides.
An unstated cell is recorded as absent and its check is skipped.

\paragraph{Conditions.} An order that exists only under a large applied field,
under pressure or in a doped sample is a different order from the ambient one;
silence about field and pressure means ambient. When one side states no vector,
a stated monolayer, few-layer flake or thin film against a stated bulk crystal
is a different order. A
transition temperature is evidence about which phase is meant: temperatures that
cannot be reconciled within 2\,K or 10\% go to the temperature judge, which
blocks the match only if they describe different phases. Silence about
temperature blocks nothing.

\paragraph{Moments.} A moment that is zero on one side and not on the other, or
that differs in size by more than half on the same site, is a different order. Moments that lie in the basal plane on one side and along the
unique axis on the other are a different order, except for rotating orders
(spirals) and cubic materials, where no unique axis exists. A rigid rotation of
all moments within the plane does not break a match.

\paragraph{Symmetry labels.} When both sides give a magnetic space group, or both
an irreducible representation, the labels are compared like with like and never
instead of the ordering type. Agreeing labels settle a vector that is missing or
disagrees, but never a unit-cell conflict. Code confirms two labels as the same only when
they are clearly one label in two spellings, and treats a catalog (BNS) number
as naming the group it stands for; anything doubtful goes to the adjudicator.

\paragraph{Applied by the scorer only.} Three provisions were added to the scorer
after answers had been read, and they are not in the text the model saw. A ferromagnetic order that
states no propagation vector is read as $\kvec=(0,0,0)$; the filled-in vector
can confirm the other side's vector but never contradict it, and the judges are
always shown the order as written. When one material's answers and key
orders can be paired in more than one way, the pairing that matches the most
orders is used, with ties broken in favor of the superseded key order and the
earliest-listed answer. And two unit cells are compared, both by the match rule and by the cell level of the detail score, only when both orders state a propagation vector, since without a vector the cell has nothing to define (Appendix~\ref{app:postrun}).

\section{The judges in full}
\label{app:judged}

\paragraph{Why these comparisons are judged.} For the unit cell, a fixed rule
comparing crystal system, centering and lattice parameters rejected correct
matches, and a simpler check, whether both sides name the same space-group
symbol, would not generalize: two descriptions can use one cell under different
symbols, or share a centering letter while one cell is a supercell of the other.
For the temperature, a mechanical fix (ignore a conflict when both sides name the
same kind of sample) gave the same scores as ignoring every conflict, because the
key orders where conflicts arise never name a sample. Code cannot tell that a
few-layer sample ordering near 45\,K is a different system from a bulk crystal
ordering near 61\,K; reading the text can.

\paragraph{Construction.} Every judge uses \texttt{claude-haiku-4-5} at
temperature 0 and sees only the fields its question concerns. The cell,
temperature and condition judges and the adjudicator take three votes decided
by majority; the citation judge makes one call per citation and claim and was checked only after it had been used (below). A failed call leaves the fixed rule's decision
standing and is never saved. A reply that cannot be read is still an answer: for
the condition judge it leaves the keyword decision standing, and for the other
judges it is a vote against the match. The voting judges' verdicts are
saved, keyed on what the judge was shown, the model, the vote count, the question
and the prompt version, and the citation judge's on the citation, the claim and
the prompt version, so a verdict is replayed only for the exact question it answered.

\paragraph{Validation before use.} The cell and temperature judges were added
after an earlier experiment had been read. Before any score was computed with
them, we fixed the answers they must give where crystallography or the sample
settles it: settings of one rhombohedrally centered lattice are the same cell, a
monoclinic cell against a hexagonal parent is not, and any pair whose text names
a monolayer, few-layer or thin-film sample, a distinct polymorph or an applied
pressure describes different phases. On the 70 distinct questions of that
experiment, 48 had such a fixed answer and the judges gave it on all 48. One
inconsistency was found: two near-identical disputed temperatures on
Ba$_2$CeIrO$_6$ received opposite verdicts. The condition judge was checked on
17 cases with documented answers, including five where keywords had misread a
sentence, and gave all 17. The adjudicator was checked on 123 pairs with known
answers (97 built automatically, such as an order against itself or against
itself with its vector's sign flipped or changed, 6 real cases from earlier experiments,
and 20 pairs of symmetry labels in different notations), and gave 121;
of its two misses, one is a pair the fixed rule rejects before the adjudicator is
asked and the other a hedged symmetry label that no longer reaches it. All four
checks were repeated on 2026-10-18 with every vote a fresh call, because the
model gateway we use had been returning cached replies, so the original three
votes had been one answer repeated. Every verdict came back the same, and none
of the adjudicator's 123 questions split its three votes (the other judges save
only the majority, so their splits were not measured).

\paragraph{On this experiment's own conflicts.} Every judge verdict the reported scores
needed had been saved, so none was missing. This experiment's answers raise
\val{n_axis_verdicts} distinct cell and temperature questions,
including some that an earlier check settles before a judge is reached, and on every one that has a fixed answer the judges gave it except two (a FePSe$_3$ unit cell and a Na$_2$Co$_2$TeO$_6$ phase), neither of which changes a score. The condition judge's verdicts and the adjudicator's were likewise all
replayed from saved verdicts, with none missing.

\paragraph{The citation judge.} This judge was at first asked whether an abstract supports a claim without being told the material, and it then marked as contradicted some claims whose abstract discussed a second material, or read a high-pressure study as contradicting an ambient claim. It is now told the material, which changed many verdicts, most of them toward ``supported''. To tell a correction from a lenient judge, we labeled 31 of the changed and unchanged pairs by hand from what the judge saw. The judge as first written was wrong on 19 of them, and the judge we use on 8, in both directions: it credited a specific pattern, such as zigzag or A-type, that the abstract states only in coarser terms, and it made two calls that were too strict. On cases like these, about one verdict in four is wrong, and ``supported'' gives too much credit to claims about a specific pattern.

\section{The three conditions and the rule that keeps them comparable}
\label{app:conditions}

A \emph{confound} here is any difference between the conditions other than the
evidence they are given. The project's rule, in substance, is:

\begin{quote}
The only things that may vary across conditions are (1) what evidence reaches
the model and (2) what it is told about that evidence. The model, the sampling
parameters, the answer-token cap, the output contract (the answer fields every
condition must fill), the instructions for weighing evidence, the match
definition, the confidence rubric and the scorer are held identical. A
difference in any of those is a confound rather than a condition.
\end{quote}

During development the rule was broken twice, and each time a measured
conclusion changed without any error. In the first, the full agent was required
to report the temperature an order applies to while the other two conditions
were not; since the scorer skips the temperature check when a field is empty,
omitting it could only raise their scores. In the second, the two simpler
conditions had not been given the match definition or the confidence rubric, so
the comparison partly measured which condition had been told the scoring rules.
The rule is now enforced in code: every condition's prompt is built from one
file whose tool instructions appear only for conditions that have those tools,
one function defines the answer fields, and a test checks that the shared parts
of every prompt are identical character for character, that every condition
receives the match definition and every tier of the confidence rubric, and that no
condition without tools is told to call one. The one part of the rubric given
only to the full agent is the rule that excludes potentially predatory journals and retracted papers,
because only a condition with the checking tools can act on it.
The prompts are too long to print here (about 3{,}900 words for closed-book and 4{,}400 for the full agent). Every condition's prompt is rendered from \texttt{agent/loop\_prompt.md} and \texttt{agent/prompt.md} in the code repository; the judges' prompts are in \texttt{agent/scorer/adjudicator.py} (cell, temperature and adjudicator), \texttt{agent/scorer/condition\_judge.py} and \texttt{agent/eval/citation\_faithfulness.py}.

Abstracts are never shortened in any condition, because an abstract often states
the refined order in its last sentence. The single search's one query is
sent without the two filters the full agent's search applies (the potentially predatory journal exclusion
and the English-only filter), so a paper from a potentially predatory journal can reach its prompt; although
neither simpler condition is told of the exclusion, citing such a paper sets
that order's citation score to $0$, as in every condition.

\section{Answer-key provenance}
\label{app:groundtruth}

Table~\ref{tab:key} lists the answer key's orders and the papers they come from.
Every value in the key is stored with the file and page of the passage that
states it, and that record is published with the code. An automated test checks that no value in the key is taken from a \magndata{} entry file. The propagation vector
$(0,0,0)$ of the three ferromagnets follows from the definition and no paper is
cited for it. The vectors of Ba$_2$LaRuO$_6$, Ba$_2$LuReO$_6$ and the second NiI$_2$ order
are left empty because no source states them in a form we could use. The vector of La$_2$NiO$_4$, $(\tfrac12,\tfrac12,0)$, is inferred: its paper gives only the in-plane components $(\tfrac12,\tfrac12)$, in the tetragonal cell. No paper we collected states a unit cell for \val{n_no_cell} of the materials, and \val{n_keys_with_msg} orders carry a magnetic space group. The propagation vectors of four materials come from papers supplied by a co-author, who later confirmed the values.

\begin{table}[t]
\centering
\caption{The answer key: one row per order, with its sources. The sources are every paper the key cites for any field of the order, so a paper may support only one field, such as the unit cell or the transition temperature.}
\label{tab:key}
\small
\begin{tabular}{>{\raggedright\arraybackslash}p{0.16\linewidth}>{\raggedright\arraybackslash}p{0.35\linewidth}>{\raggedright\arraybackslash}p{0.16\linewidth}>{\raggedright\arraybackslash}p{0.22\linewidth}}
\toprule
material & ordering type (status) & $\mathbf{k}$ & sources \\
\midrule
Ba$_2$CeIrO$_6$ & A-type antiferromagnetic & $(0,0,1)$ & \citep{aczel2019,revelli2019} \\
\rowrule
Ba$_2$IrO$_4$ & basal-plane antiferromagnetic, moments along $[110]$ & $(\tfrac12,\tfrac12,0)$ & \citep{boseggia2013,clancy2023} \\
\rowrule
Ba$_2$LaRuO$_6$ & Type IIIa antiferromagnetic & empty & \citep{battle1983,battle1989} \\
\rowrule
Ba$_2$LuReO$_6$ & Type I antiferromagnetic & empty & \citep{xiong2018} \\
\rowrule
Ba$_2$YReO$_6$ & antiferromagnetic & $(0,0,1)$ & \citep{nilsen2021} \\
\rowrule
CoPS$_3$ & zigzag antiferromagnetic, small moment component along $c$ (current) & $(0,1,0)$ & \citep{wildes2017} \\
 & collinear zigzag antiferromagnetic, moments along $a$ (superseded) & $(0,1,0)$ & \citep{wildes2017} \\
\rowrule
CrGeTe$_3$ & ferromagnetic & $(0,0,0)$ & \citep{zhu2021} \\
\rowrule
CrI$_3$ & ferromagnetic & $(0,0,0)$ & \citep{mcguire2015,ke2021} \\
\rowrule
CrSBr & A-type antiferromagnetic & $(0,0,\tfrac12)$ & \citep{scheie2022} \\
\rowrule
CrSiTe$_3$ & ferromagnetic & $(0,0,0)$ & \citep{zhu2021} \\
\rowrule
FePS$_3$ & zigzag antiferromagnetic & $(0,1,\tfrac12)$ & \citep{coak2021,mccreary2020,amirabbasi2023} \\
\rowrule
FePSe$_3$ & zigzag antiferromagnetic & $(\tfrac12,0,\tfrac12)$ & \citep{chen2024,basnet2024} \\
\rowrule
La$_2$NiO$_4$ & two-sublattice N\'eel antiferromagnetic & $(\tfrac12,\tfrac12,0)$, inferred & \citep{petsch2023} \\
\rowrule
MnTe & A-type antiferromagnetic & $(0,0,0)$ & \citep{mazin2023,amin2024} \\
\rowrule
Na$_2$Co$_2$TeO$_6$ & triple-$\mathbf{k}$ antiferromagnetic (current) & $(\tfrac12,0,0)$, three arms & \citep{chen2021,samarakoon2021} \\
 & single-$\mathbf{k}$ zigzag antiferromagnetic (superseded) & $(\tfrac12,0,0)$ & \citep{chen2021,samarakoon2021} \\
\rowrule
NiI$_2$ & incommensurate helical, below 60\,K (current) & $(0.1384,0,1.457)$ & \citep{kuindersma1981,liu2025nii2} \\
 & collinear antiferromagnetic, 60 to 73\,K (current) & empty & \citep{kuindersma1981,liu2025nii2} \\
\rowrule
UO$_2$ & transverse triple-$\mathbf{k}$ antiferromagnetic (current) & $[0,0,1]$ & \citep{paddison2025} \\
 & single-$\mathbf{k}$ antiferromagnetic (superseded) & $[0,0,1]$ & \citep{paddison2025} \\
\rowrule
YbBi$_2$IO$_4$ & stripe antiferromagnetic & $(\tfrac12,0,0)$ & \citep{park2025} \\
\bottomrule
\end{tabular}
\end{table}

\section{Score definitions and the pre-registration}
\label{app:score}

\begin{table}[t]
\centering
\caption{The total score and its four parts.}
\label{tab:weights}
\begin{tabular}{llr}
\toprule
part & what it measures & weight \\
\midrule
correctness (confidence-weighted $F_1$) & whether the answer is right & 45\% \\
\rowrule
detail score & whether it is detailed enough to act on & 25\% \\
\rowrule
top-1 accuracy & whether the most confident order is a current one & 15\% \\
\rowrule
citation integrity & whether the cited papers hold up & 15\% \\
\bottomrule
\end{tabular}
\end{table}

\paragraph{Correctness.} 
For each material, every reported order and key order that pass the match rule
are candidate pairs, and the pairing with the most matches is chosen
(Appendix~\ref{app:match}). A matched order earns credit $1$, a type-only match
$0.5$, and a match that states no unit cell where the key gives one, or a cell
that cannot be compared with the key's, $0.75$; when either order states no propagation vector the cell is not compared and earns $1$. Over all materials together, with $T$ the total credit and
$N$ the number of key orders, recall is $T/N$ and precision is $T/(T+\sum c)$.
The sum runs over unmatched reported orders, each of which adds a penalty $c$ set
by its stated confidence $p$:
$c=0.6\,[1-\cos(\pi p/160)]$ for $p\le80$ and $c=0.6+0.02\,(p-80)$ above,
halved if the model marked the order contested and $1$ if it gave no confidence.
An unmatched order adds no penalty when it would match a key order that another
reported order already matched, or when it describes the material under other
conditions and the citation judge found a cited abstract that supports it
(Appendix~\ref{app:postrun}). Correctness is $2PR/(P+R)$. A plain $F_1$ that
counts every unmatched order as $1$ is computed alongside and published.

\paragraph{Detail score and top-1.} The detail score averages, over all matched
pairs, the levels of Appendix~\ref{app:vocabulary}, weighted $10$, $8$, $4$ and
$3$ for type, vector, cell and symmetry; a level the key does not state is left
out and the others are scaled up. Top-1 is $1$ when the order the model gave the
highest confidence matches a current key order, and $0$ otherwise.

\paragraph{Citation integrity.} A citation whose DOI resolves earns $1.0$, which drops to $0.8$ when the judge finds that the abstract does not address the claim. A paper found only by its title earns $0.5$, which becomes $0.6$ if its abstract supports the claim and $0.4$ if it does not address it. A record with no abstract is not judged, so the citation keeps the credit for existing. A citation that leads to no indexed paper earns $0$. An order takes the credit of
its best citation, except that a citation to a paper that does not exist (a DOI
that does not resolve and whose title matches no indexed paper), is ineligible
or is contradicted by its abstract sets the order's credit to $0$; a title alone
that matches no indexed paper earns $0$ without setting the order's credit to $0$.
An order with no citation scores $0$, and an order whose citations could not be
resolved at all is left out rather than scored $0$.

\paragraph{Missing parts.} When a part of the score cannot be computed for a
repeat, its weight is shared among the other parts rather than scored zero, so a
failure to match anything is not punished three times (on correctness, top-1 and
detail).

\paragraph{The pre-registration.} After the first three materials had been run
in an early version of the experiment, and before any other, we wrote down the
weights and three reporting rules: the total score is never reported without its
four parts, a part that cannot be computed shares its weight rather than scoring
zero, and an incomplete evaluation is labeled as incomplete. On the smaller set
we had then, the ranking depended on the weights (closed-book and the full agent
each came first under about $45\%$ of weightings). We also designed two further
parts and left them out of the total: one that rewarded reporting a superseded
order with a lower confidence than the current one, because the right reward depends on why the user
asked, and a calibration part, because a coarse answer at moderate confidence can
be well calibrated while giving little detail. Both are reported separately.
Whether the total score agrees with how a person would judge the answers has not
been tested; the planned check is for a person to rank a sample of answers
without seeing their scores.

\paragraph{Robustness to the weights.} We recomputed the total score under every
set of four weights that are multiples of $5\%$ and sum to $100\%$,
\val{simplex_grid_points} weightings in all. The full agent comes first under
every one of them, because its mean is the highest of the three conditions on
each of the four parts: correctness \val{s7_cwf1}, detail score
\val{s7_specificity}, top-1 \val{s7_top1} and citation integrity
\val{s7_citations}, against at most \val{s1_cwf1}, \val{s0_specificity},
\val{s1_top1} and \val{s1_citations} for the other two. Its lead on the detail score over closed-book is within the uncertainty (\ci{gap_s7_s0_specificity}), so under weightings that put most of the weight on detail, the two are not clearly separated. The dependence on the
weights found at pre-registration, on the smaller early set, does not appear in
this experiment.
\begin{table}[t]
\centering
\caption{The detail score and two of its levels. A level earns credit only where
the order matched and its value is correct, and a level the key does not state
is skipped rather than scored zero.}
\label{tab:specificity}
\input{tables/specificity}
\end{table}

\paragraph{Adding a criterion.} The four detail levels suit the question we
asked, and other questions need others: a transition temperature, the
temperature range of each phase, or whether an order belongs to the bulk or a
monolayer. A criterion here is three things declared together in one file: a
field the agent is asked to fill, a rule for comparing two values of it, and a
weight. A field declared without a rule is reported but not graded. Every
criterion must say when the field should be left empty, because a field asked
for without that condition invites a guessed value. In this experiment every condition
reported a transition temperature, each phase's temperature range and the
dimensionality of the sample, at zero weight.

\section{The magnetic-order vocabulary and the detail-score tiers}
\label{app:vocabulary}

\begin{figure}[tp]
\centering
\includegraphics[width=0.958\linewidth]{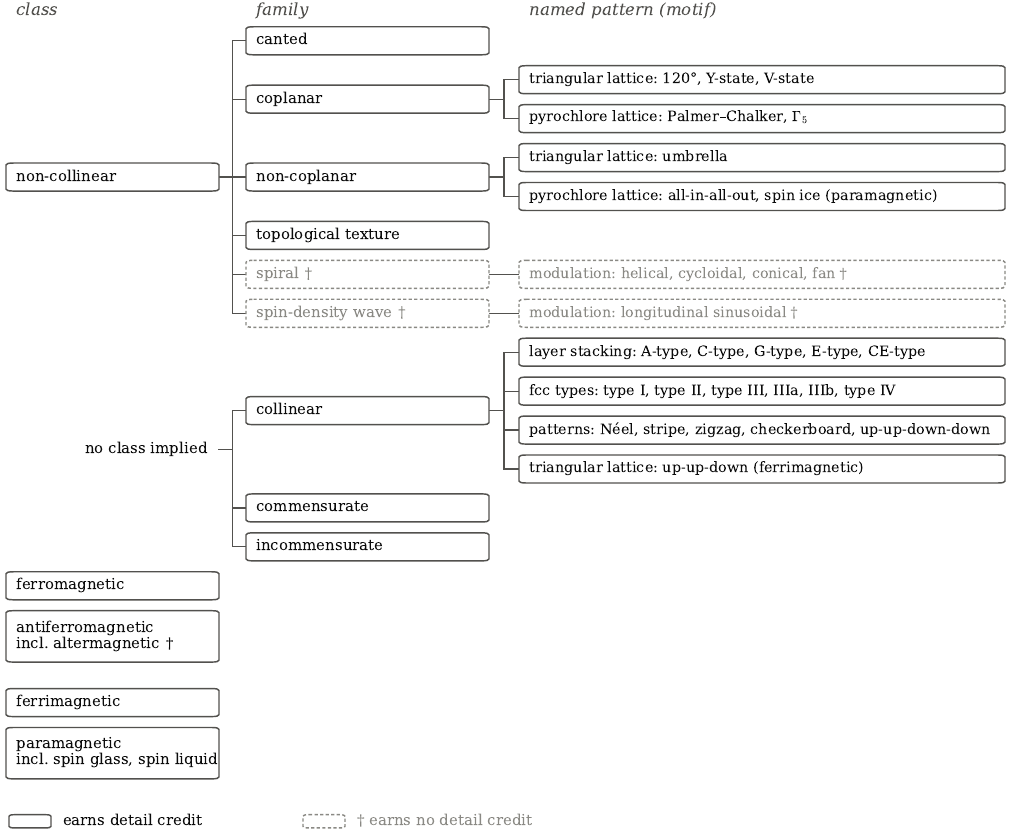}
\caption{The ordering-type vocabulary the detail score uses, as the scorer defines
it. A line joins a term to the broader term it belongs to; collinear,
commensurate and incommensurate imply no class. Every named pattern is
antiferromagnetic unless its class is given in brackets, and the modulations are
non-collinear. Topological textures are skyrmions and related textures; types
IIIa and IIIb are finer forms of type III. Terms marked \dag{} earn no detail
credit, most of them because the match rule already compares them; the terms for
the number of propagation vectors and for the moment direction, not shown, earn
none either.}
\label{fig:app-vocabulary}
\end{figure}

Which of two ordering-type names is more specific cannot be read off the
strings, so the scorer uses a vocabulary of terms at four levels: class
(ferromagnetic, antiferromagnetic, ferrimagnetic, non-collinear), family
(collinear, coplanar, incommensurate), motif (zigzag, A-type, N\'eel, stripe) and
direction. The full list is generated from
the scorer and published with the code. \textbf{Figure~\ref{fig:app-vocabulary}} shows the
classes, families and named patterns and how they nest: each named pattern
belongs to one family, and a family belongs to at most one class.

A term earns detail credit only if the match rule does not already decide it
through the propagation vector, the modulation type or the number of vectors, so
these are not counted twice. Specific patterns such as zigzag, A-type and N\'eel
earn credit, because the match rule checks the propagation vector rather than
whether the pattern was named. The modulation type (helical, cycloidal,
longitudinal sinusoidal) and the number of propagation vectors (single- against
multi-\kvec{}) do not, because the match rule compares them already, and neither
does the moment direction. Nor can such a term earn its credit back through a
broader name: crediting ``helical order'' for belonging to the family of
rotating-moment orders would give back the credit just withheld, so that family
earns no detail credit.

The ordering-type, propagation-vector and unit-cell levels score $1$, $0.5$ or
$0$ against what the key states. Half credit goes to an answer exactly one level
less precise than the key and consistent with it: the whole set of
symmetry-equivalent vectors where the key names one, a space-group symbol where
the key also gives lattice parameters, or the right family, such as ``collinear'',
where the key names a pattern such as zigzag. A bare class such as
``antiferromagnetic'' earns half credit only against a key whose finest term is a
canted, coplanar or non-coplanar family, which all belong to the non-collinear
class; against a key that names a pattern, or a collinear, commensurate or
incommensurate family, it earns nothing. A pattern that contradicts the
key's (G-type where the key says A-type) scores zero. The symmetry level scores
$1$ for the key's label, $0.5$ when one side lists alternatives that include the
other side's single label, and $0$ otherwise. In this experiment the key states a
symmetry label for seven orders: an irreducible representation for Ba$_2$IrO$_4$,
Ba$_2$YReO$_6$, La$_2$NiO$_4$ and both CoPS$_3$ orders, and a magnetic space group for YbBi$_2$IO$_4$
and Na$_2$Co$_2$TeO$_6$'s superseded order.

\section{The citation check in full}
\label{app:citations}

The first step asks separately whether a citation's identifier resolves and
whether the paper exists, so an unresolved identifier is not counted as a
fabrication. The search is by DOI and, separately, by title, because the
closed-book condition often cites a paper by its title, or by author and year,
without a DOI. Titles are looked up in Crossref and then in OpenAlex, because
Crossref indexes older work poorly. Every lookup is saved, filed under the DOI or title
looked up, so re-scoring needs no network access. The second step judges every
citation whose paper was found, by DOI or by title, against the abstract in the
paper's bibliographic record. In this experiment the answers gave \val{n_unique_citations} distinct
citations. Of these, \val{n_unique_citations_unresolved} were given only by a title whose closest indexed match was too similar to reject and too different to accept, and are left out rather than scored. A paper with no abstract in its record is not judged, and keeps the credit for existing.

\begin{table}[t]
\centering
\caption{Citation integrity by condition. ``Skipped'' counts orders none of
whose citations could be resolved; they are left out, not scored 0.}
\label{tab:citations}
\input{tables/citations}
\end{table}

\section{Reproducibility}
\label{app:repro}
\label{sec:repro}

Every table and every number taken from the generated numbers file is written by
a script in the code repository from the saved answers, with every judge replayed
from its saved verdicts, so regenerating them needs no network access and no
model calls. The agent's answers themselves cannot be reproduced by running it
again, because the model is not deterministic. The closed-book answers come from an earlier experiment; we checked that every request that condition sends, from the system prompt onward, is identical under the code of that experiment and of this one, and reused them, so only the single search and the full agent were run again. No run of this experiment was set aside or run again, and every run used one version of the code. The standard deviation we report is taken over the repeat totals, and which repeat of one material is grouped with which repeat of another is arbitrary, so it is a rough guide rather than a confidence interval. When the single search was re-run on three materials during development, with the same retrieval and the scorer of that time, its mean correctness on them moved from $0.685$ to $0.643$, more than the standard deviation within each run ($0.031$ and $0.021$), so we treat a difference below about $0.05$ as noise. The single-search and full-agent runs used about 53 million
model tokens, and the agent's offline test suite (2{,}475 tests under Python 3.9)
passes at the revision that regenerates these numbers.

\paragraph{Confidence intervals.} The intervals in the Results are 95\% percentile bootstrap intervals. Each of \val{ci_resamples} resamples draws \val{n_headline} materials with replacement, keeps all \val{n_repeats} repeats of each, and recomputes every number the way the paper computes it, with the same scorer and the same saved judge verdicts: the total score, its four parts, the staleness score and the calibration error per repeat and then the mean over repeats, and the percentages and the mean confidence of confident orders over all repeats together. We resample materials rather than repeats because the materials vary far more than the repeats do, so an interval shows how much a number could change on a different set of materials. The same resample is used for all three conditions, so the interval on a difference between two conditions is a paired one. The full agent's expected calibration error lies near the bottom of its interval: the error adds up absolute gaps between confidence and accuracy, so the noise that resampling adds tends to raise it. The staleness score exists only for the \val{n_stale_pairs} materials with a superseded order, so its interval is taken over the resamples that contain at least one of them. The intervals in \textbf{Figure~\ref{fig:permaterial}} instead resample the \val{n_repeats} repeats of each material, since a single material has nothing else to resample.

\textbf{Table~\ref{tab:ci}} gives the interval for every number in the Results that has one.

\begin{table}[tp]
\centering
\caption{The numbers in the Results with their 95\% confidence intervals, from
resampling the \val{n_headline} materials. (a) Each condition. (b) Differences
between conditions.}
\label{tab:ci}
\footnotesize
\begin{tabular}{>{\raggedright\arraybackslash}p{0.30\linewidth}lll}
\toprule
(a) & \Szero{} & \Sone{} & \Sseven{} \\
\midrule
total score & \val{s0_composite}~\ci{s0_composite} & \val{s1_composite}~\ci{s1_composite} & \val{s7_composite}~\ci{s7_composite} \\
\rowrule
correctness & \val{s0_cwf1}~\ci{s0_cwf1} & \val{s1_cwf1}~\ci{s1_cwf1} & \val{s7_cwf1}~\ci{s7_cwf1} \\
\rowrule
detail score & \val{s0_specificity}~\ci{s0_specificity} & \val{s1_specificity}~\ci{s1_specificity} & \val{s7_specificity}~\ci{s7_specificity} \\
\rowrule
top-1 & \val{s0_top1}~\ci{s0_top1} & \val{s1_top1}~\ci{s1_top1} & \val{s7_top1}~\ci{s7_top1} \\
\rowrule
citation integrity & \val{s0_citations}~\ci{s0_citations} & \val{s1_citations}~\ci{s1_citations} & \val{s7_citations}~\ci{s7_citations} \\
\rowrule
orders with no propagation vector & \val{s0_no_k_pct}\%~\ci{s0_no_k_pct} & \val{s1_no_k_pct}\%~\ci{s1_no_k_pct} & \val{s7_no_k_pct}\%~\ci{s7_no_k_pct} \\
\rowrule
checkable orders whose cited titles lead to no paper & \val{s0_orders_unfound_pct}\%~\ci{s0_orders_unfound_pct} & \val{s1_orders_unfound_pct}\%~\ci{s1_orders_unfound_pct} & \val{s7_orders_unfound_pct}\%~\ci{s7_orders_unfound_pct} \\
\rowrule
cited abstracts that contradict the claim & \val{s0_contradicted_pct}\%~\ci{s0_contradicted_pct} & \val{s1_contradicted_pct}\%~\ci{s1_contradicted_pct} & \val{s7_contradicted_pct}\%~\ci{s7_contradicted_pct} \\
\rowrule
current against superseded (staleness score) & \val{s0_staleness}~\ci{s0_staleness} & \val{s1_staleness}~\ci{s1_staleness} & \val{s7_staleness}~\ci{s7_staleness} \\
\rowrule
expected calibration error & \val{s0_ece}~\ci{s0_ece} & \val{s1_ece}~\ci{s1_ece} & \val{s7_ece}~\ci{s7_ece} \\
\rowrule
mean confidence, orders at confidence $\ge 80$ & \val{s0_hiconf_conf}~\ci{s0_hiconf_conf} & \val{s1_hiconf_conf}~\ci{s1_hiconf_conf} & \val{s7_hiconf_conf}~\ci{s7_hiconf_conf} \\
\rowrule
orders at confidence $\ge 80$ that match the key & \val{s0_hiconf_correct_pct}\%~\ci{s0_hiconf_correct_pct} & \val{s1_hiconf_correct_pct}\%~\ci{s1_hiconf_correct_pct} & \val{s7_hiconf_correct_pct}\%~\ci{s7_hiconf_correct_pct} \\
\bottomrule
\end{tabular}

\vspace{6pt}
\begin{tabular}{>{\raggedright\arraybackslash}p{0.30\linewidth}ll}
\toprule
(b) & total score & correctness \\
\midrule
\Soneup{} over \Szero{} & \val{gap_s1_s0_composite}~\ci{gap_s1_s0_composite} & \val{gap_s1_s0_cwf1}~\ci{gap_s1_s0_cwf1} \\
\rowrule
\Ssevenup{} over \Sone{} & \val{gap_s7_s1_composite}~\ci{gap_s7_s1_composite} & \val{gap_s7_s1_cwf1}~\ci{gap_s7_s1_cwf1} \\
\rowrule
\Ssevenup{} over \Szero{} & \val{gap_s7_s0_composite}~\ci{gap_s7_s0_composite} & \val{gap_s7_s0_cwf1}~\ci{gap_s7_s0_cwf1} \\
\bottomrule
\end{tabular}

\vspace{6pt}
\begin{tabular}{>{\raggedright\arraybackslash}p{0.30\linewidth}lll}
\toprule
(b, continued) & \Soneup{} over \Szero{} & \Ssevenup{} over \Sone{} & \Ssevenup{} over \Szero{} \\
\midrule
detail score & \ci{gap_s1_s0_specificity} & \ci{gap_s7_s1_specificity} & \ci{gap_s7_s0_specificity} \\
\rowrule
top-1 & \ci{gap_s1_s0_top1} & \ci{gap_s7_s1_top1} & \ci{gap_s7_s0_top1} \\
\rowrule
citation integrity & \ci{gap_s1_s0_citations} & \ci{gap_s7_s1_citations} & \ci{gap_s7_s0_citations} \\
\rowrule
orders with no propagation vector & \ci{gap_s1_s0_no_k_pct} & \ci{gap_s7_s1_no_k_pct} & \ci{gap_s7_s0_no_k_pct} \\
\rowrule
cited abstracts that contradict the claim & \ci{gap_s1_s0_contradicted_pct} & \ci{gap_s7_s1_contradicted_pct} & \ci{gap_s7_s0_contradicted_pct} \\
\rowrule
expected calibration error & \ci{gap_s1_s0_ece} & \ci{gap_s7_s1_ece} & \ci{gap_s7_s0_ece} \\
\rowrule
orders at confidence $\ge 80$ that match the key & \ci{gap_s1_s0_hiconf_correct_pct} & \ci{gap_s7_s1_hiconf_correct_pct} & \ci{gap_s7_s0_hiconf_correct_pct} \\
\bottomrule
\end{tabular}
\end{table}

\section{Per-material outcomes and the match-rule audit}
\label{app:permaterial}

Figure~\ref{fig:permaterial} reports correctness rather than a per-material
total, because per material the detail score exists only where an order matched,
so a per-material total would combine different parts for different materials.

\begin{figure}[tp]
\centering
\includegraphics[width=0.847\linewidth]{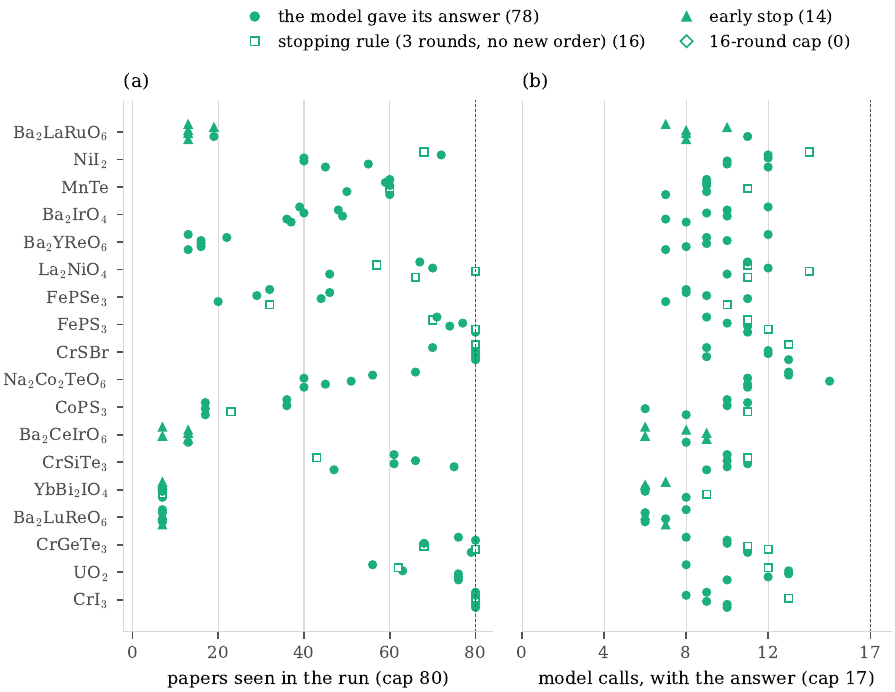}
\caption{Every full-agent run, one marker per run (\val{n_repeats} per material), with its shape showing what ended the run. (a) Papers the run's searches returned. (b) Model calls the run made, counting the call that gave the answer, so a run ended by the 16-round cap shows 17. Materials are in the order of \textbf{Figure~\ref{fig:permaterial}}.}
\label{fig:app-runs}
\end{figure}

\paragraph{How the full agent's runs ended.} The literature the searches find splits the materials roughly into two groups (\textbf{Figure~\ref{fig:app-runs}}). On Ba$_2$CeIrO$_6$, Ba$_2$LaRuO$_6$, Ba$_2$LuReO$_6$, Ba$_2$YReO$_6$ and YbBi$_2$IO$_4$, no run saw more than 22 papers, and all \val{s7_early_stops} early stops fall on these materials. On the other 13 materials every run saw at least 36 papers, except on CoPS$_3$ and FePSe$_3$ (17 to 46), and \val{s7_paper_cap_runs} of the \val{s7_stop_total} runs reached the 80-paper cap, after which the agent kept reading and comparing until the stopping rule, the round cap or its own answer ended the run.

\begin{figure}[tp]
\centering
\includegraphics[width=0.488\linewidth]{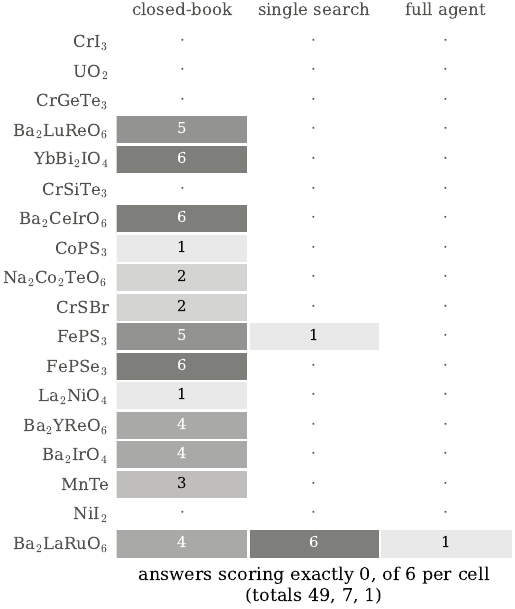}
\caption{Answers that score exactly zero, out of \val{n_repeats} per material and condition; a dot marks none. Materials are in the order of \textbf{Figure~\ref{fig:permaterial}}.}
\label{fig:app-zeros}
\end{figure}

\paragraph{Where the zero scores fall.} Closed-book's \val{s0_zeros} answers that score exactly zero are spread over 13 of the \val{n_headline} materials, and on Ba$_2$CeIrO$_6$, FePSe$_3$ and YbBi$_2$IO$_4$ every repeat scores zero (\textbf{Figure~\ref{fig:app-zeros}}). The single search's \val{s1_zeros} fall on 2 materials, 6 of them on Ba$_2$LaRuO$_6$, and the full agent's \val{s7_zeros} on Ba$_2$LaRuO$_6$. All \val{s0_abstentions} of closed-book's abstentions fall on YbBi$_2$IO$_4$, and all \val{s1_abstentions} of the single search's on Ba$_2$LaRuO$_6$. On FePS$_3$ and FePSe$_3$, closed-book names a zigzag antiferromagnetic order in every repeat, and every propagation vector it gives is $(0,1,0)$, stated at confidence $55$ to $90$, where the key has $(0,1,\tfrac12)$ and $(\tfrac12,0,\tfrac12)$ respectively.

\paragraph{Three materials shaped by the stopping rule or the key.} The early stop of \S\ref{sec:agent} has a cost and a benefit, and both show on single materials. It fired on \val{s7_early_stops} runs over \val{s7_early_stop_materials} materials. On Ba$_2$LaRuO$_6$ it ended \val{s7_early_stops_laru} of the \val{n_repeats} runs: the searches find few papers on the material (at most 19 in a run), the two oldest from the early 1980s, one of which is the source of our key \citep{battle1983}, so further searching had little left to find. On Ba$_2$CeIrO$_6$ it ended \val{s7_early_stops_ceir} of the \val{n_repeats} runs, often before the neutron study that determines the order \citep{aczel2019} turned up. That study is cited in none of the \val{n_repeats} full-agent answers, against \val{aczel_cited_earlier} in an earlier experiment with a version of the agent that had no early stop, which found it through searches that did not name the material. On Ba$_2$YReO$_6$ the full agent loses top-1 in \val{yre_top1_lost} of \val{yre_repeats} repeats because its most confident order cites a 2025 study that places the N\'eel temperature at $31\,$K, with a quadrupolar phase above it \citep{omar2025}, while our key, taken from an earlier neutron study \citep{nilsen2021}, records transitions at $45$ and $20\,$K; the temperature judge treats the two as different phases. Here the agent follows newer literature than our key.

\paragraph{Current against superseded orders, material by material.} The
staleness score of \S\ref{sec:results-staleness} is $1$ when the current order is
given a higher confidence than the superseded one and $0.5$ when only the current
order is reported. On UO$_2$ every repeat of every condition scores at least
$0.5$; the single search scores $1$ in all \val{n_repeats} repeats, the full agent
in 4 and closed-book in 2. On CoPS$_3$ the full agent scores $1$ in two repeats and
$0.5$ in the other four, closed-book scores $0.5$ in four repeats and $0$ in the other two,
and the single search scores $0.5$ in one repeat and $0$ in the rest. On Na$_2$Co$_2$TeO$_6$ the single
search scores $1$ in every repeat and the full agent in 2 (and $0.5$ in one more); every
other repeat scores $0$. There, the current triple-\kvec{} and the superseded single-\kvec{} descriptions are both reported
and both marked contested in all but one repeat of the single search and of the full
agent, which reflects a live dispute in the literature as much as an error;
closed-book never reports the current order.

\begin{figure}[tp]
\centering
\includegraphics[width=0.967\linewidth]{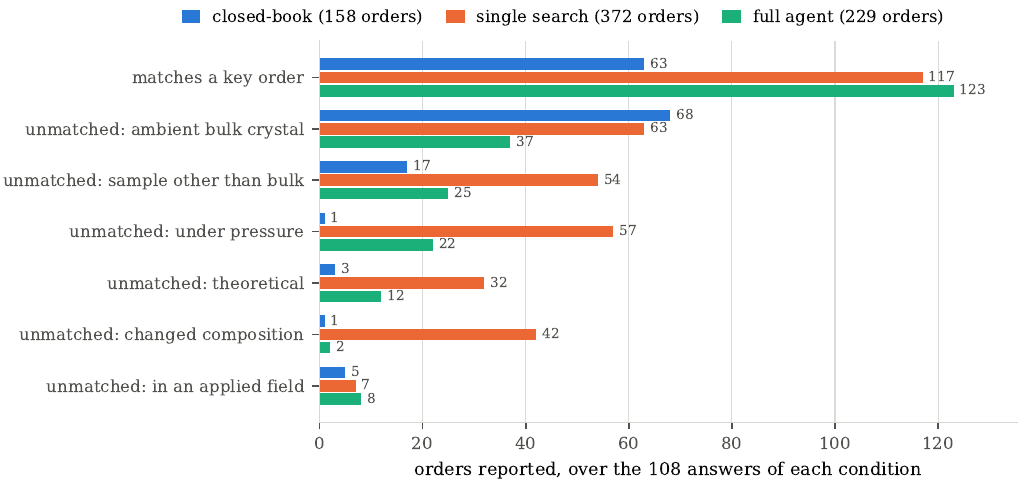}
\caption{Every order each condition reported, over all of its answers: those that match a key order, and those that match none, by what they describe. An unmatched order that restates an order already matched in the same answer is counted under what it describes.}
\label{fig:app-orders}
\end{figure}

\paragraph{What the unmatched orders are.} \textbf{Figure~\ref{fig:app-orders}} sorts every reported order by what it describes. Every reported order is tagged by the
first of these rules that applies: a theoretical method named in the order or its
conditions; a sample other than bulk; or doping, high pressure or an applied
field. An order none of them fires on is counted as the ambient bulk crystal.
Of the unmatched orders, \val{s0_unmatched_dup}, \val{s1_unmatched_dup} and
\val{s7_unmatched_dup} (closed-book, single search, full agent) restate an order
that another order of the same answer already matched, and \val{s0_unmatched_offscope},
\val{s1_unmatched_offscope} and \val{s7_unmatched_offscope} describe something
other than the ambient bulk crystal; an order can be both. The full agent reports orders other than the ambient bulk crystal on 14 of the \val{n_headline} materials. After the rule of
Appendix~\ref{app:postrun}, the penalties that remain sum to
\val{s0_fp_cost}, \val{s1_fp_cost} and \val{s7_fp_cost} over all materials and
repeats. Every order, its tags and its penalty are published with the paper's
sources.

\paragraph{The per-criterion audit.} Each check that can reject a match outright
was switched off in turn, with the judges off, and all \val{n_comparisons}
comparisons of a reported order with a key order re-scored
(Table~\ref{tab:matchrule}a). Without the fixed unit-cell rule, the total score would rise by \val{gate_unit_cell_delta_s0}, \val{gate_unit_cell_delta_s1}
and \val{gate_unit_cell_delta_s7}, and without the temperature rule by
\val{gate_temperature_delta_s0}, \val{gate_temperature_delta_s1} and
\val{gate_temperature_delta_s7}; these are the two comparisons now sent to
judges. Panel (b) shows the scores with neither judge, either one or both.

\begin{table}[tp]
\centering
\caption{What each part of the match rule does to the scores.}
\label{tab:matchrule}
\input{tables/matchrule}
\end{table}

\section{Limitations in detail}
\label{app:limitations}

\paragraph{The scorer.} The detail score rewards stating a correct value, but a
temperature, field or pressure conflict can reject a match only when an order
states one, so on these criteria the scorer favors an answer that leaves its
conditions out. Keywords decide which orders the condition judge reads, so an
order whose wording no keyword catches is treated as ambient without a judge
reading it.

\paragraph{What the comparison separates.} We did not record how much literature
exists for each material, so we cannot say how much of closed-book's spread
across materials that explains. Repeats of one run also vary less than separate runs do (Appendix~\ref{app:repro}).

\section{Scoring decisions made after the experiment}
\label{app:postrun}

The scorer was changed after the answers had been read, each time for every
condition alike and on the saved answers, so nothing was run again. Most of the changes correct how the scorer reads what an answer wrote: a notation for a propagation vector, a symmetry label or a unit cell that it misread, a denial such as ``no long-range order'' that it took for a claim, a moment direction stated inside a longer clause. Each was found by reading answers, and every answer each one changes was read. Several changes are decisions about what the score should reward rather than corrections, chief among them:
reading a ferromagnet's blank propagation vector as $(0,0,0)$; using the pairing that matches the most
orders instead of pairing first-come, first-served; comparing two unit cells only when both orders state a propagation vector; and the rule that an unmatched order costs nothing when it
restates a matched order, or when it describes the material under conditions the
key does not cover and a cited abstract supports it. We wrote that last rule down and
committed it before computing its effect. One change concerns the citation judge, which is now told which material the claim is about, so that a paper about a related material can no longer support it. The last five changes were decided after this experiment's full-agent answers had been read. Table~\ref{tab:postrun} gives the
total score and correctness as the changes are applied one after another. The
last row is the reported result, and the first, the score with none of the changes, already ranks the conditions in the same order. The full agent ran after every change but the last five had been made, and its compare tool used that scorer, so for that condition the first row is a counterfactual rather than the score it was given at the time.

The cell and temperature judges were also added to the match rule after an earlier experiment had been read; without them the totals would be \val{s0_composite_prejudge}, \val{s1_composite_prejudge} and \val{s7_composite_prejudge}, in the same order.

\begin{table}[tp]
\centering
\caption{Total score (correctness in brackets) as the scoring changes made
after the answers were read are applied cumulatively. A row that groups several changes gives the score after the whole group; the \texttt{RESULTS.md} file of the code repository lists every change on its own row.}
\label{tab:postrun}
\footnotesize
\input{tables/postrun}
\end{table}

\section{The same experiment on other agent models}
\label{app:models}

\begin{figure}[p]
\centering
\includegraphics[width=0.6\linewidth,height=0.4\textheight,keepaspectratio]{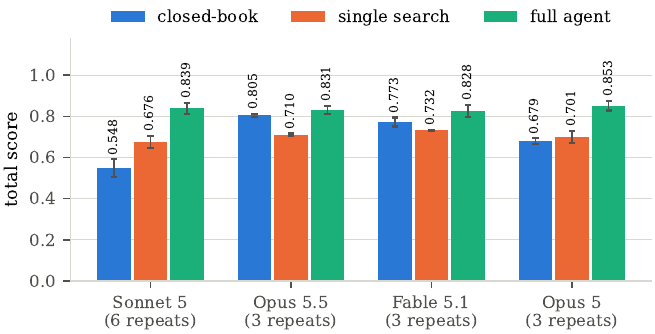}
\caption{The total scores of \textbf{Table~\ref{tab:models}}; whiskers are one standard deviation across repeats.}
\label{fig:app-models}

\vspace{1.3em}

\includegraphics[width=0.9\linewidth,height=0.77\textheight,keepaspectratio]{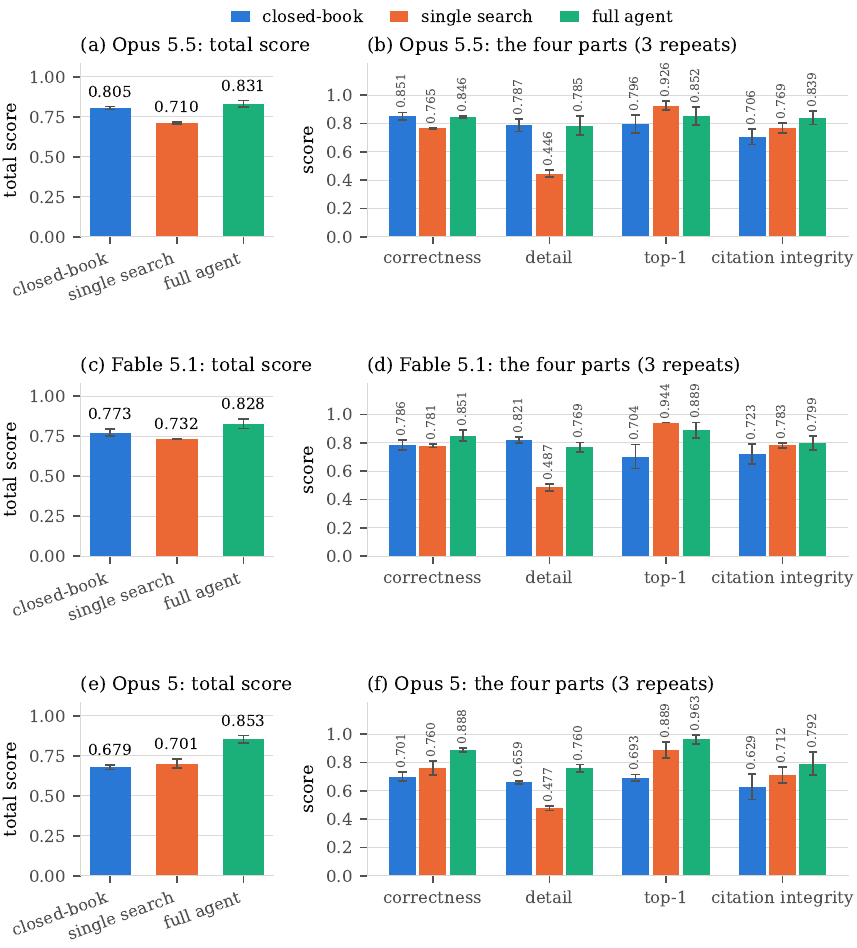}
\caption{Figure~\ref{fig:conditions}a and b for the three other agent models: for each model, the total score (left) and its four parts (right), means over 3 repeats with whiskers one standard deviation across repeats.}
\label{fig:app-models-parts}
\end{figure}

To see whether the comparison depends on the agent model, we ran the same experiment on three other models: \texttt{claude-opus-5-5}, \texttt{claude-fable-5-1} and \texttt{claude-opus-5}. All four models ran in one experiment under the same conditions: the single search and the full agent were run for every model on the same code and from the same saved judge verdicts and paper summaries, so no model's run could use another's, and the closed-book answers were reused, because their requests are identical and use no literature. Seven full-agent repeats of the other models, which had lost access to arXiv, were run again on the same agent code. Only the agent model differs; the prompt, tools, caps, answer key, scorer and judges (\texttt{claude-haiku-4-5}) are the same. \texttt{claude-sonnet-5} ran \val{n_repeats} repeats, the main results of this paper, and each other model ran 3, so their standard deviations rest on fewer repeats (\textbf{Table~\ref{tab:models}}, \textbf{Figure~\ref{fig:app-models}}).

The full agent has the highest mean total score with every model, and its lead over the single search holds with every model: the 95\% interval on the gap, from resampling the materials as in Appendix~\ref{app:repro}, excludes zero. Its lead over closed-book is clear only with \texttt{claude-sonnet-5} and \texttt{claude-opus-5}; with \texttt{claude-opus-5-5} and \texttt{claude-fable-5-1}, the interval includes zero. With these two models, closed-book's mean is also above the single search's, because these models give the key's propagation vector from memory on materials where \texttt{claude-sonnet-5} does not (for example FePS$_3$ and FePSe$_3$), but this difference is also within the uncertainty. \textbf{Figure~\ref{fig:app-models-parts}} shows the four parts for each of the three other models: as with \texttt{claude-sonnet-5}, the single search gives the least detail of the three conditions with every model.

\begin{table}[h]
\centering
\caption{Total score of each condition for each agent model, mean over repeats $\pm$ one standard deviation across repeats, scored with the same scorer and saved judge verdicts as the main results.}
\label{tab:models}
\footnotesize
\input{tables/models}
\end{table}

%% file: tables/specificity.tex
\begin{tabular}{lrrr}
\toprule
condition & detail score & $\mathbf{k}$ level & unit-cell level \\
\midrule
\Szero{} & 0.721 \,$\pm$\,0.020 & 0.872 \,$\pm$\,0.076 & 1.000 \,$\pm$\,0.000 \\
\rowrule
\Sone{} & 0.411 \,$\pm$\,0.044 & 0.417 \,$\pm$\,0.049 & 0.283 \,$\pm$\,0.157 \\
\rowrule
\Sseven{} & 0.730 \,$\pm$\,0.047 & 0.822 \,$\pm$\,0.062 & 0.798 \,$\pm$\,0.094 \\
\midrule
\multicolumn{4}{@{}p{\dimexpr\linewidth-2\tabcolsep}@{}}{\footnotesize Weights within the total score: type 10\%, k 8\%, cell 4\%, symmetry 3\%. A level for which the key has no value is skipped, not scored zero.} \\
\bottomrule
\end{tabular}

%% file: tables/citations.tex
\begin{tabular}{lrrr}
\toprule
condition & citation integrity & orders checked & skipped \\
\midrule
\Szero{} & 0.425 \,$\pm$\,0.102 & 151 & 7 \\
\rowrule
\Sone{} & 0.812 \,$\pm$\,0.035 & 372 & 0 \\
\rowrule
\Sseven{} & 0.867 \,$\pm$\,0.020 & 229 & 0 \\
\midrule
\multicolumn{4}{@{}p{\dimexpr\linewidth-2\tabcolsep}@{}}{\footnotesize Checks applied: full (the paper exists, is not excluded, and its abstract supports the claim). A skipped check is left out of the mean, not scored as a failure. Citation integrity is averaged over the orders in each repeat, then over the repeats, as in the total score; averaging over all orders at once instead gives a slightly different value (\Szero{} 0.4253 here vs 0.4265 pooled).} \\
\bottomrule
\end{tabular}

%% file: tables/matchrule.tex
\begin{tabular}{llrrr}
\toprule
\multicolumn{5}{l}{\textit{(a) each rule that rejects a match, switched off one at a time}} \\
\addlinespace[2pt]
rule rejects on & pairs it rejected first & \Szero{} & \Sone{} & \Sseven{} \\
\midrule
moment direction & 24 & +0.017 & -0.001 & -0.008 \\
\rowrule
\textbf{unit cell} & 9 & +0.005 & +0.000 & +0.001 \\
\rowrule
\textbf{temperature} & 40 & +0.002 & +0.000 & +0.012 \\
\rowrule
order type & 255 & +0.002 & +0.023 & +0.005 \\
\rowrule
single-k vs multi-k & 80 & +0.000 & +0.000 & +0.000 \\
\rowrule
number of k arms & 2 & +0.000 & +0.000 & +0.000 \\
\rowrule
moment size & 0 & +0.000 & +0.000 & +0.000 \\
\rowrule
applied field / pressure & 93 & +0.000 & +0.001 & +0.002 \\
\addlinespace[4pt]
\multicolumn{5}{l}{\textit{(b) sending the unit-cell and temperature questions to a judge, one, the other, or both}} \\
\addlinespace[2pt]
question sent to a judge & part of score & \Szero{} & \Sone{} & \Sseven{} \\
\midrule
neither (fixed rule only) & total score & 0.543 & 0.676 & 0.833 \\
 & correctness & 0.530 & 0.743 & 0.863 \\
 & detail & 0.717 & 0.411 & 0.733 \\
 & top-1 & 0.407 & 0.778 & 0.870 \\
\addlinespace[2pt]
unit cell only & total score & 0.546 & 0.676 & 0.833 \\
 & correctness & 0.532 & 0.743 & 0.863 \\
 & detail & 0.724 & 0.411 & 0.733 \\
 & top-1 & 0.417 & 0.778 & 0.870 \\
\addlinespace[2pt]
temperature only & total score & 0.544 & 0.676 & 0.839 \\
 & correctness & 0.534 & 0.743 & 0.874 \\
 & detail & 0.714 & 0.411 & 0.730 \\
 & top-1 & 0.407 & 0.778 & 0.889 \\
\addlinespace[2pt]
both (reported) & total score & \textbf{0.548} & \textbf{0.676} & \textbf{0.839} \\
 & correctness & 0.536 & 0.743 & 0.874 \\
 & detail & 0.721 & 0.411 & 0.730 \\
 & top-1 & 0.417 & 0.778 & 0.889 \\
\midrule
\multicolumn{5}{@{}p{\dimexpr\linewidth-2\tabcolsep}@{}}{\footnotesize Panel (a): how much each condition's total score changes when that rule is switched off, with both judges off, over all 975 comparisons of a reported order with a key order. A positive value means the fixed rule was rejecting answers that would otherwise score; a negative value means it was rejecting answers that should not. The two rules in bold are the ones now sent to a judge. Panel (b): the total score and its parts when the unit-cell question, the temperature question, or both are sent to a judge instead of decided by the fixed rule.} \\
\bottomrule
\end{tabular}

%% file: tables/postrun.tex
\begin{tabular}{p{0.52\linewidth}ccc}
\toprule
change (cumulative) & closed-book & single search & full agent \\
\midrule
scored with none of the changes below & 0.529 (0.560) & 0.567 (0.519) & 0.712 (0.695) \\
\rowrule
+ pairing that matches the most orders (was first-come) & 0.528 (0.568) & 0.567 (0.519) & 0.721 (0.708) \\
\rowrule
+ a ferromagnet's blank \kvec{} reads as $(0,0,0)$ & 0.528 (0.568) & 0.606 (0.569) & 0.748 (0.739) \\
\rowrule
+ 4 further fixes to how the scorer reads answers & 0.528 (0.568) & 0.609 (0.574) & 0.780 (0.767) \\
\rowrule
+ restatements, and supported orders under other conditions, cost nothing & 0.536 (0.585) & 0.625 (0.610) & 0.783 (0.774) \\
\rowrule
+ 23 further fixes to how the scorer reads answers & 0.533 (0.541) & 0.624 (0.587) & 0.802 (0.789) \\
\rowrule
+ the citation judge is told which material the claim is about & 0.543 (0.541) & 0.652 (0.698) & 0.800 (0.829) \\
\rowrule
+ 5 further fixes to how the scorer reads answers & 0.536 (0.531) & 0.652 (0.698) & 0.812 (0.839) \\
\rowrule
+ the unit cell is compared only when both orders state a propagation vector & 0.543 (0.531) & 0.663 (0.723) & 0.825 (0.851) \\
\rowrule
+ 7 further fixes to how the scorer reads answers & 0.546 (0.531) & 0.659 (0.716) & 0.825 (0.851) \\
\rowrule
+ the five changes decided after this experiment's full-agent answers had been read & 0.548 (0.536) & 0.676 (0.743) & 0.839 (0.874) \\
\bottomrule
\end{tabular}

%% file: tables/models.tex
\begin{tabular}{lcccc}
\toprule
agent model & repeats & \Szero{} & \Sone{} & \Sseven{} \\
\midrule
Sonnet 5 (main results) & 6 & 0.548\,$\pm$\,0.044 & 0.676\,$\pm$\,0.031 & 0.839\,$\pm$\,0.027 \\
Opus 5.5 & 3 & 0.805\,$\pm$\,0.009 & 0.710\,$\pm$\,0.007 & 0.831\,$\pm$\,0.020 \\
Fable 5.1 & 3 & 0.773\,$\pm$\,0.022 & 0.732\,$\pm$\,0.003 & 0.828\,$\pm$\,0.029 \\
Opus 5 & 3 & 0.679\,$\pm$\,0.015 & 0.701\,$\pm$\,0.029 & 0.853\,$\pm$\,0.024 \\
\bottomrule
\end{tabular}